\documentclass[12pt]{article}
\usepackage{amssymb,amsmath}

\usepackage{curves}
\usepackage{epic}
\usepackage{eepic}
\usepackage{epsfig}

\newcommand{\dd}{\mbox{\rm d}}

\newcommand{\p}{\partial}

\newcommand{\be}{\begin{equation}}
\newcommand{\bear}{\begin{eqnarray}}

\newcommand{\ear}{\end{eqnarray}}
\newcommand{\ee}{\end{equation}}
\newcommand{\lbl}{\label}
\newcommand{\bi}{\bibitem}
\newcommand{\ci}{\cite}

\newcommand{\vs}{\vspace}
\newcommand{\hs}{\hspace}

\begin{document}

\begin{center}

\
\vs{0.6cm}

\baselineskip .7cm

{\bf \Large Stable Self-Interacting Pais-Uhlenbeck Oscillator}

\vs{2mm}

\baselineskip .5cm
Matej Pav\v si\v c

Jo\v zef Stefan Institute, Jamova 39, SI-1000, Ljubljana, Slovenia; 

email: matej.pavsic@ijs.si

\vs{3mm}

{\bf Abstract}
\end{center}

\baselineskip .43cm
{\small

It is shown that the interacting Pais-Uhlenbeck oscillator necessarily
leads to a description with a Hamiltonian that contains positive and
negative energies associated with two oscillators. Descriptions with a
positive definite Hamiltonians, considered by some authors, can hold
only for a free Pais-Uhlenbeck oscillator. We demonstrate that the solutions
of a self-interacting Pais-Uhlenbeck oscillator are stable on islands
in the parameter space, as already observed in the literature. If we
slightly modify the system, by considering a sine interaction term,
and/or by taking unequal masses of the two oscillators, then the system
is stable on the continents that extend from zero to infinity in the parameter
space. Therefore, the Pais-Uhlenbeck oscillator is quite acceptable
physical system.
}

\baselineskip .55cm

\section{Introduction}

The ordinary gravity, described by the Einstein-Hilbert action (containing the
curvature scalar $R$), is not renormalizable. The higher derivative gravity,
with $R+R^2$, is renormalizable. But higher derivative theories are considered
as problematic, because according the the Ostrogradski
formalism\,\ci{Ostrogradski} they contain
negative energies which, according to the wide spread belief, automatically
imply instabilities at the classical and quantum level. It is often stated
that at the quantum level such a theory implies negative probabilities due to
ghosts, and is therefore not unitary. Whether a theory implies negative
probabilities and positive energies, or vice versa, positive probabilities
and negative energies, depends on choice of vacuum and corresponding
creation and annihilations operators\,\ci{Pauli}--\ci{Ozonder}.

A model for a higher derivative theory is the Pais-Uhlenbeck
oscillator\,\ci{PU}. It has been studied by many authors,
because understanding the issues
concerning its stability, could pave the way towards quantum gravity.
Smilga\,\ci{Smilga1}--\ci{SmilgaStable} has found that there are islands of
stability of the classical solutions of the interacting
Pais-Uhlenbck (PU) oscillator. An example of an unconditionally stable
interacting system was also
found\,\ci{SmilgaStable}. This system, which is a non linear extension of the
PU oscillator, is a close relative of a
supersymmetric higher-derivative system\,\ci{Robert}. Recently,
Mostafazadeh\,\ci{Mostafazadeh} has  found a Hamiltonian formulation
of the PU oscillator that yields a stable and unitary quantum system.
Other authors\,\ci{Bolonek}--\ci{Nucci} have also arrived at the positive
definite Hamiltonians for the PU oscillator. A procedure
with a PT symmetric Hamiltonian without ghosts and negative energies
in the spectrum has been considered in Refs.\,\ci{Bender}--\ci{Mannheim:2012ci}.
In this paper we show
that the descriptions of Ref.\,\ci{Mostafazadeh}--\ci{Nucci} hold for a free
PU oscillator only, but not for
a self-interacting one. In the latter case one has to describe the PU oscillator
by the second order Lagrangian and employ the Ostrogradski formalism.
Then, as it is well known, the PU oscillator can be
written as a system of two coupled oscillators, one with positive and the
other one with negative energy. Stability issues arise, because the
energy can flow from one to the other oscillator in such a way that
their kinetic energies escape into positive and negative infinity, respectively,
while the total energy of the system remains constant and finite.

In this paper we first show explicitly that, in general, a self-interacting
PU oscillator
cannot be described by a positive definite Hamiltonian. Then we consider
numerical solutions of the PU oscillator with the quartic self-interaction.
Unless the coupling constant or the initial velocity
is too high, the (classical) system is stable. So we indeed have islands
of stability as observed by Smilga\,\ci{Smilga1}--\ci{SmilgaStable},
and recently by\,\ci{Ilhan}. Then we consider two
modifications of the PU oscillator. (i) We replace the quartic
interaction term with a term that contains the forth power of sine.
We show numerically
and analytically that such a modification gives infinite continents of stability.
(ii) Instead of taking equal masses of the two oscillators, we consider
the case in which the masses are different. Such a modified system is in
fact just like the PU oscillators, only the coefficients in front
of the terms are changed, and a non-linear term is added. Again we obtain
stability for the vast range of parameters. Moreover, regardless of how
high the initial velocity is, the solution is stable. Such behavior
of classical solution implies that the quantum system is stable
as well\,\ci{Ilhan}.

\section{The Pais-Uhlenbeck oscillator as a system of two oscillators}

The Pais-Uhlenbeck oscillator is given by the 4th order differential equation
\be
   \left ( \frac{\dd^2}{\dd t^2} + \omega_1^2 \right ) 
   \left (\frac{\dd^2}{\dd t^2} + \omega_2^2 \right ) x = 0,
\lbl{2.1}
\ee
which gives
\be
   x^{(4)} + (\omega_1^2 + \omega_2^2) {\ddot x} + \omega_1^2 \omega_2^2 x = 0.
\lbl{2.2}
\ee
As observed by Mostafazadeh\,\ci{Mostafazadeh}, this can be written as
the system of two oscillators\footnote{We use here a different notation for
coefficients.}
\be
   {\ddot x} + \mu_1 x - \rho_1 y = 0,
\lbl{2.3}
\ee
\be
    {\ddot y} + \mu_2 y - \rho_2 x = 0,
\lbl{2.4}
\ee
where $\mu_1,~\mu_2,~\rho_1,~\rho_2$ are real constants.

From Eq.\,(\ref{2.3}) we have $y=(1/\rho_1)({\ddot x} +\mu_1 x)$. Inserting this
into Eq.\,(\ref{2.4}), we obtain
\be
    x^{(4)} + (\mu_1 + \mu_2) {\ddot x} +(\mu_1 \mu_2 - \rho_1 \rho_2) x = 0.
\lbl{2.5}
\ee
Comparison of the latter equation with (\ref{2.2}) gives the relations
\be
     \mu_1 +\mu_2 = \omega_1^2 +\omega_2^2
\lbl{2.6}
\ee
\be
    \mu_1 \mu_2 - \rho_1 \rho_2 = \omega_1^2 \omega_2^2 .
\lbl{2.7}
\ee                 
The solutions is
\be
    \omega_{1,2}^2 = \mbox{$\frac{1}{2}$} (\mu_1 +\mu_2) \pm \mbox{$\frac{1}{2}$} 
    \sqrt{(\mu_1 +\mu_2)^2 - 4 (\mu_1 \mu_2 - \rho_1 \rho_2)} .
\lbl{2.8}
\ee

Let us now find possible Lagrangians corresponding to the equations of
motion (\ref{2.3}),(\ref{2.4}).

{\bf Case I.}

Assuming the Lagrangian
\be
    L=\mbox{$\frac{1}{2}$}({\dot x}^2 + {\dot y}^2) - \mbox{$\frac{1}{2}$}
    (\mu_1 x^2 + \mu_2 y^2 - 2 \rho_1 x y ),
\lbl{2.9}
\ee
we obtain the equations of motion (\ref{2.3}),(\ref{2.4}), if
\be
    \rho_2 = \rho_1 .
\lbl{2.10}
\ee
Then from Eq.\,(\ref{2.8}) we have
\be
    \omega_{1,2}^2 = \mbox{$\frac{1}{2}$}(\mu_1 + \mu_2) \pm \mbox{$\frac{1}{2}$} 
    \sqrt{(\mu_1 -\mu_2)^2 + 4  \rho_1^2} .
\lbl{2.11}
\ee
We see that for a large range of the coefficients $\mu_1$, $\mu_2$, $\rho_1$,
the squared frequencies $\omega_1^2$ and $\omega_2^2$ are positive. Then
$\omega_1$, $\omega_2$ are real, in which case we have oscillating motion.

The Hamiltonian is
\be
    H= \mbox{$\frac{1}{2}$}(p_x^2 + p_y^2) + \mbox{$\frac{1}{2}$}(\mu_1 x^2
+\mu_2 y^2 - 2 \rho_1 x y),
\lbl{2.12}
\ee
where $p_x=\p L/\p {\dot x} = {\dot x}$, and $p_y =\p L/\p {\dot y} = {\dot y}$.
By performing a rotation in the $(x,y)$-space,
\bear
    &&~x'=x\, {\rm cos}\, \alpha + y\, {\rm sin}\, \alpha \nonumber\\
    && y' = -x\, {\rm sin}\, \alpha + y\, {\rm cos}\, \alpha  
\lbl{2.13}
\ear
with the accompanying rotation of momenta,
\bear
    &&~p_{x'}=p_x\, {\rm cos}\, \alpha + p_y\, {\rm sin}\, \alpha \nonumber\\
    && p_{y'}= -p_x\, {\rm sin}\, \alpha + p_y\, {\rm cos}\, \alpha,  
\lbl{2.14}
\ear  
the Hamiltonian (\ref{2.12}) can be diagonalized. By comparing the new Hamiltonian
\be
      H= \mbox{$\frac{1}{2}$}(p_{x'}^2 + p_{y'}^2) + \mbox{$\frac{1}{2}$}(a {x'}^2
         +b {y'}^2),
\lbl{2.15}
\ee  
with the old one (\ref{2.12}), we obtain the system of three equations for
the unknowns $a$, $b$, $\alpha$:
\bear
     && a\, {\rm cos}^2 \,\alpha + b\, {\rm sin}^2 \, \alpha = \mu_1 \nonumber\\
     && a\, {\rm sin}^2 \,\alpha + b\, {\rm cos}^2 \, \alpha = \mu_2 \lbl{2.16}\\
     && (a-b)\, {\rm cos}\, \alpha \,{\rm sin}\, \alpha = \rho_1 \nonumber .
\ear
The solution is
\be
    a = \mbox{$\frac{1}{2}$}(\mu_1 + \mu_2) + \mbox{$\frac{1}{2}$} 
    \sqrt{(\mu_1 -\mu_2)^2 + 4  \rho_1^2} = \omega_1^2.
\lbl{2.17}
\ee
\be
    b = \mbox{$\frac{1}{2}$}(\mu_1 + \mu_2) - \mbox{$\frac{1}{2}$} 
    \sqrt{(\mu_1 -\mu_2)^2 + 4  \rho_1^2} = \omega_2^2.
\lbl{2.18}
\ee
\be
     {\rm cos}\, 2 \alpha= \frac{\mu_1-\mu_2}{\sqrt{(\mu_1 -\mu_2)^2 + 4  \rho_1^2}}.
\lbl{2.19}
\ee
The $a$, $b$ are just equal to the squared frequencies $\omega_1^2$, $\omega_2^2$
of the PU oscillator. This can be directly verified by inserting the expressions
(\ref{2.16}) into the equations of motion (\ref{2.5}) and using (\ref{2.10}).

In the new coordinates, the system is described by the Lagrangian\footnote{The
author\,\ci{Mostafazadeh} has also arrived at such a system of two uncoupled oscillators
straightforwardly from Eq.\,(\ref{2.2}), by using a different chain of
substitutions of variables. See also the procedure of refs.\,\ci{Nucci}.}
\be
    L=\mbox{$\frac{1}{2}$}({\dot x}'^2 + {\dot y}'^2) - \mbox{$\frac{1}{2}$}
    (\omega_1^2 {x'}^2 + \omega_2^2 {y'}^2),
\lbl{2.20}
\ee
and the Hamiltonian
\be
    H=\mbox{$\frac{1}{2}$}({\dot x}'^{\,2} + {\dot y}'^{\,2}) + \mbox{$\frac{1}{2}$}
    (\omega_1^2 {x'}^2 + \omega_2^2 {y'}^2).
\lbl{2.21}
\ee
The energy of this system is positive. It is remarkable that when we diagonalize
the $L$ and $H$ for a system of two oscillators (\ref{2.3}) and (\ref{2.4}),
we obtain two different frequencies, $\omega_1$ and $\omega_2$, that correspond
to those occurring in the PU oscillators.

{\bf Case II.}

Alternatively, we may assume that the Lagrangian is
\be
    L=\mbox{$\frac{1}{2}$}({\dot x}^2 - {\dot y}^2) - \mbox{$\frac{1}{2}$}
    (\mu_1 x^2 - \mu_2 y^2 - 2 \rho_1 x y ),
\lbl{2.22}
\ee
This gives the equations of motion (\ref{2.3}) and (\ref{2.4}) if
\be
     \rho_2 = - \rho_1 \,.
\lbl{2.23}
\ee
Inserting this into Eq.\,(\ref{2.8}), we obtain
\be
    \omega_{1,2}^2 = \mbox{$\frac{1}{2}$}(\mu_1 + \mu_2) \mp \mbox{$\frac{1}{2}$} 
    \sqrt{(\mu_1 -\mu_2)^2 - 4  \rho_1^2} .
\lbl{2.24}
\ee
The frequencies $\omega_1$, $\omega_2$ are real if
$(\mu_1 -\mu_2)^2 > 4  \rho_1^2$ and $\mu_1 + \mu_2 >
\sqrt{(\mu_1 -\mu_2)^2 - 4  \rho_1^2}$.

The Hamiltonian is 
\be
    H= \mbox{$\frac{1}{2}$}(p_x^2 - p_y^2) + \mbox{$\frac{1}{2}$}(\mu_1 x^2
-\mu_2 y^2 - 2 \rho_1 x y).
\lbl{2.25}
\ee
By performing the hyperbolic rotation in the $(x,y)$-space,
\bear
    &&x'=x\, {\rm cosh}\, \alpha + y\, {\rm sinh}\, \alpha \nonumber\\
    && y' =x\, {\rm sinh}\, \alpha + y\, {\rm cosh}\, \alpha  
\lbl{26}
\ear
with the accompanying rotation of momenta,
\bear
    &&p_{x'}=p_x\, {\rm cosh}\, \alpha + p_y\, {\rm sin}\, \alpha \nonumber\\
    && p_{y'}= p_x\, {\rm sinh}\, \alpha + p_y\, {\rm cosh}\, \alpha,  
\lbl{2.27}
\ear
the Lagrangian (\ref{2.22}) and the Hamiltonian (\ref{2.25}) become
\be
    L=\mbox{$\frac{1}{2}$}({\dot x}'^2 - {\dot y}'^2) - \mbox{$\frac{1}{2}$}
    (\omega_1^2 x'^2 - \omega_2^2 y'^2),
\lbl{2.28}
\ee
\be
    H=\mbox{$\frac{1}{2}$}({\dot x}'^2 - {\dot y}'^2) + \mbox{$\frac{1}{2}$}
    (\omega_1^2 x'^2 - \omega_2^2 y'^2).
\lbl{2.29}
\ee
Again, the diagonalized Lagrangian and Hamiltonian contain the frequencies
$\omega_1$, $\omega_2$ of the PU oscillator.

Now we have the relations
\bear
     && \omega_1^2\, {\rm cosh}^2 \,\alpha - \omega_2^2, {\rm sinh}^2 \, 
      \alpha = \mu_1 \lbl{2.29a}\\
     && -\omega_1^2\, {\rm sinh}^2 \,\alpha + \omega_2^2\, {\rm cosh}^2 \, \alpha = \mu_2 \lbl{2.29b}\\
     && (\omega_1^2-\omega_2^2)\, {\rm cosh}\, \alpha \,{\rm sinh}\, \alpha 
  = -\rho_1 \lbl{2.29c} .
\ear
The energy of the system is either positive or negative, depending on which degree of
freedom is more excited.

Cases I and II show that the PU oscillator can be described as a system of two
oscillators whose Hamiltonian is either (\ref{2.21}) or (\ref{2.29}).
Case I means positive definite Hamiltonian, whereas Case II means indefinite
Hamiltonian.

\section{Self-interacting PU oscillator}
\subsection{Equations of motion and the Lagrangian}

We have seen that the PU oscillator can be described as a system of two
oscillators (\ref{2.3}) and (\ref{2.4}) that can be written in the explicit
uncoupled form
\be
   {\ddot x'} +\omega_1^2 x' = 0
\lbl{3.1a}
\ee
\be
   {\ddot y'} +\omega_2^2 y' = 0
\lbl{3.2a}
\ee
For real $\omega_1$, $\omega_2$, this is an oscillating system, regardless
of whether for the corresponding Lagrangian we take (\ref{2.20})
or (\ref{2.28}). Both Lagrangians are equally good for describing the
PU oscillator\,\ci{Bolonek,Bagarello}.

If we include an interaction between the $x'$ and $y'$, then energy
can be transfered between those two degrees of freedom. Then it does
matter which Lagrangian we take.

\ (i) Let us first consider the following Lagrangian that is an extension
of (\ref{2.20}) (Case I):
\be
    L=\mbox{$\frac{1}{2}$}({\dot x}'^2 + {\dot y}'^2) - \mbox{$\frac{1}{2}$}
    (\omega_1^2 x'^2 + \omega_2^2 y'^2) - \frac{\lambda}{4} (x'+y')^4
 \lbl{3.1}
\ee
The corresponding equations of motions are
\be
   {\ddot x'} +\omega_1^2 x' + \lambda (x'+y')^3 = 0
\lbl{3.2}
\ee
\be
   {\ddot y'} +\omega_2^2 y'+ \lambda (x'+y')^3  = 0
\lbl{3.3}
\ee
Introducing the new coordinates
\be
    u=\frac{x'+y'}{\sqrt{2}}~,~~~~~~~~~~v=\frac{x'-y'}{\sqrt{2}} ,
\lbl{3.4}
\ee
we have
\be
   L=\mbox{$\frac{1}{2}$}({\dot u}^2 + {\dot v}^2) - \mbox{$\frac{1}{4}$}
    [(\omega_1^2 + \omega_2^2)(u^2+v^2) + 2 (\omega_1^2 - \omega_2^2)u v]
     - \lambda u^4
 \lbl{3.5}
\ee
\be
     {\ddot u} + \mu_1 u -\rho_1 v + 4 \lambda u^3 = 0
\lbl{3.6}
\ee
\be
     {\ddot v} + \mu_2 v -\rho_1 u = 0 ,
\lbl{3.7}
\ee
where
\be
    \mu_1 = \mu_2 = \mbox{$\frac{1}{2}$}(\omega_1^2 + \omega_2^2) ~,~~~~~
     -\rho_1 = \mbox{$\frac{1}{2}$}(\omega_1^2 - \omega_2^2).
\lbl{3.8}
\ee
Eliminating $u$, we obtain the 4th order differential equation for $v$:
\be
     v^{(4)} + (\mu_1 + \mu_2) {\ddot v} + (\mu_1 \mu_2 - \rho_1^2) v +
   4 \lambda \rho_1 ({\ddot v} + \mu_2 v)^3 =0 ,
\lbl{3.9}
\ee
which is just that of the PU oscillator with an extra non linear term.

Similarly, by eliminating $v$, we obtain 
\be
     u^{(4)} + (\mu_1 + \mu_2) {\ddot u} + (\mu_1 \mu_2 - \rho_1^2) u +
   4 \mu_2 \lambda u^3 + 4 \lambda \frac{\dd^2}{\dd t^2} \left ( u^3 \right ) = 0,
\lbl{3.10}
\ee
which is also the PU oscillator with a non-linear term.

(ii) Let us now consider the Lagrangian that is an extension of (\ref{2.28})
(Case II):
\be
    L=\mbox{$\frac{1}{2}$}({\dot x}'^2 - {\dot y}'^2) - \mbox{$\frac{1}{2}$}
    (\omega_1^2 x'^2 - \omega_2^2 y'^2) - \frac{\lambda}{4} (x'+y')^4 .
 \lbl{3.11}
\ee
The corresponding equations of motions are now
\be
   {\ddot x'} +\omega_1^2 x' + \lambda (x'+y')^3 = 0
\lbl{3.12}
\ee
\be
   {\ddot y'} +\omega_2^2 y'- \lambda (x'+y')^3  = 0
\lbl{3.13}
\ee
Notice the minus sign in the second equation.

In the new variables $u$, $v$, defined in Eq.\,(\ref{3.4}), we have
\be
   L={\dot u} {\dot v} - \mbox{$\frac{1}{4}$}
    [(\omega_1^2 - \omega_2^2)(u^2+v^2) + 2 (\omega_1^2 + \omega_2^2)u v]
     - \lambda u^4
 \lbl{3.14}
\ee
\be
     {\ddot u} + \mu_1 u -\rho_1 v = 0
\lbl{3.15}
\ee
\be
     {\ddot v} + \mu_2 v -\rho_1 u + 4 \lambda u^3= 0 ,
\lbl{3.16}
\ee
where $\mu_1,~\mu_2$ and $\rho_1$ are given in Eqs.\,(\ref{3.8}).
By eliminating $v$, we obtain
\be
    u^{(4)} + (\mu_1 + \mu_2) {\ddot u} +(\mu_1 \mu_2 -\rho_1^2) u +
  4 \rho_1 \lambda u^3 = 0.
\lbl{3.17}
\ee
Using (\ref{3.8}) and introducing $\Lambda=2 (\omega_1^2-\omega_2^2) \lambda$,
the latter equation reads
\be
    u^{(4)} + (\omega_1^2 + \omega_2^2) {\ddot u} +\omega_1 \omega_2^2 u 
    -\Lambda u^3 = 0.
\lbl{3.18}
\ee
Now we obtain the equation of motion for the PU oscillator with a self-interaction
term.
The second order Lagrangian that gives the fourth order equation of motion
(\ref{3.18}) is that of the PU oscillator with a quartic self-interaction term:
\be
    L=\mbox{$\frac{1}{2}$} \left [ {\ddot u}^2 -(\omega_1^2 + \omega_2^2) 
    {\ddot u}^2 +\omega_1 \omega_2^2 u^2 \right ]
    +\mbox{$\frac{1}{4}$} \Lambda u^4 .
\lbl{3.19}
\ee
Notice that if $\omega_1=\omega_2$, then $\rho_1=0$, and $v$ in (\ref{3.15})
cannot be expressed in terms of $u$. Consequently, in the case $\omega_1=\omega_2$,
the system (\ref{3.15}),(\ref{3.16}) does not give the equation (\ref{3.17}) for
the PU oscillator.

The Ostrogradski second order formalism then leads us to
the phase space Lagrangian\,\ci{Bolonek:2006ir,Mannheim:2000ka,Mannheim:2004qz}
that is equivalent to (\ref{3.19}), 
\be
    L= p_u {\dot u} + p_q {\dot q} - H,
\lbl{3.20}
\ee
where
\be
   H = p_u q +\mbox{$\frac{1}{2}$} \left [ p_q^2 + (\omega_1^2 + \omega_2^2)q^2
- \omega_1^2 \omega_2^2 u^2 \right ] + \mbox{$\frac{1}{4}$} \Lambda u^4 .
\lbl{3.21}
\ee
The latter Hamiltonian can be transformed into
\be
   H'= \mbox{$\frac{1}{2}$}(p_{x'}^2 - p_{y'}^2) + \mbox{$\frac{1}{2}$}
      (\omega_1^2 x'^2 - \omega_2^2 y'^2) + \frac{\lambda}{4} (x'+y')^4 .
\lbl{3.24}
\ee
The phase space Lagrangian
\be
   L' = p_{x'} {\dot x'} + p_{y'} {\dot y'} - H'
\lbl{3.25}
\ee
is equivalent to (\ref{3.11}). This can be directly seen by using the equations
of motion $p_{x'}={\dot x'}$ and $p_{y'}={\dot y'}$, and eliminating
$p_{x'}$, $p_{y'}$ from (\ref{3.25}).

Therefore,
the correct procedure is to start from the 4th order self-interacting
Lagrangian (\ref{3.19}) and to employ the Ostrogradski formalism. The Hamiltonian
so obtained can be positive or negative. The procedures discussed
in Refs.\,\ci{Mostafazadeh}, and also in Refs.\,\ci{Bolonek}--\ci{Nucci},
have limited validity, because they do not consider an interaction
term. They are valid descriptions of the PU oscillator in the absence of an
interaction, but not if one switches on an interaction.

\subsection{Solutions}

The Lagrangian of the form (\ref{3.11}) is usually considered as unsuitable
for physics, because it implies indefinite Hamiltonian, with
positive and negative energy states. The interaction term that mixes
the two types of states leads to instabilities. But as pointed out
in Refs.\,\ci{Smilga1}--\ci{SmilgaStable},\ci{Ilhan}, there exist islands of stability. We show this
explicitly by solving numerically the equations of motion
(\ref{3.12}), (\ref{3.13}). In Fig.\,1 there are examples of such
calculations, done by MATHEMATICA. In all examples we take $\omega_1^2=1$ and $\omega_2^2=1.5$

\setlength{\unitlength}{.8mm}

\begin{figure}[h!]
\hs{3mm}
\begin{picture}(120,120)(25,0)

\put(25,61){\includegraphics[scale=0.4]{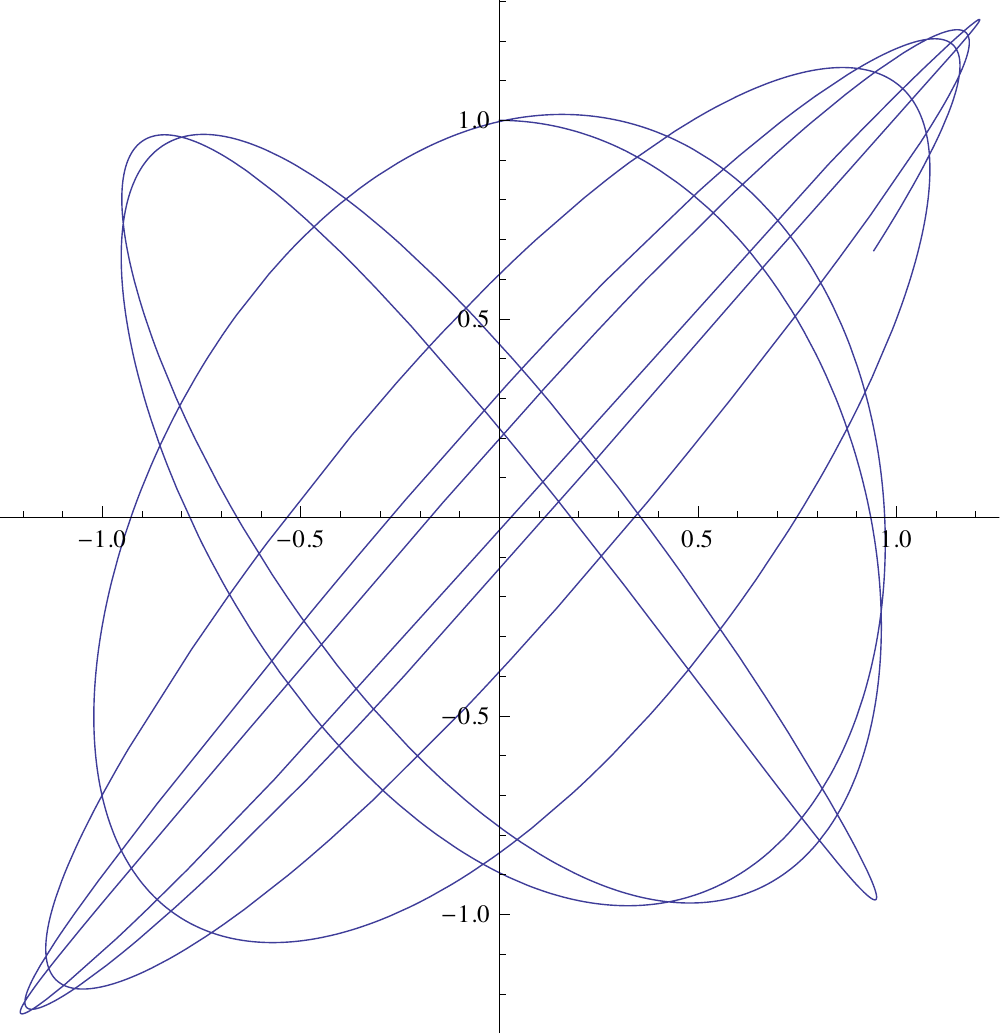}}
\put(90,61){\includegraphics[scale=0.4]{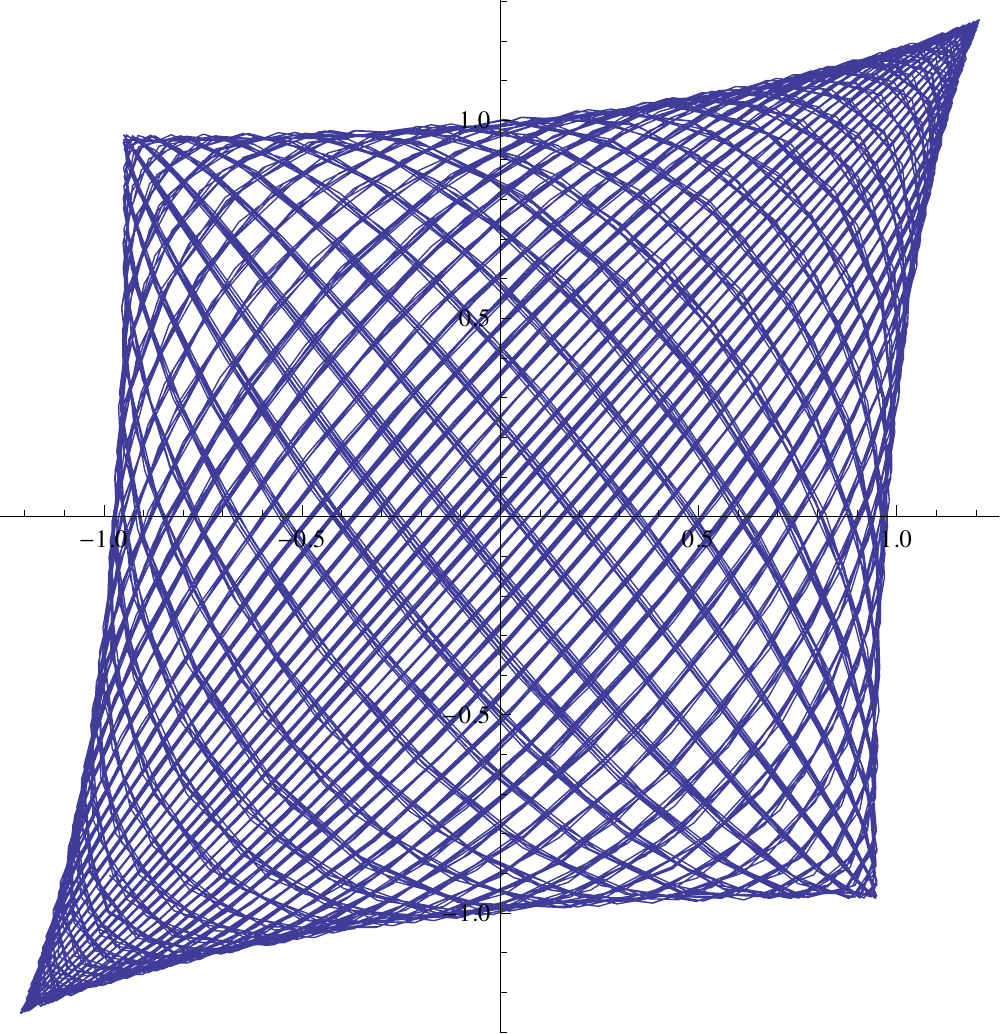}}
\put(150,61){\includegraphics[scale=0.4]{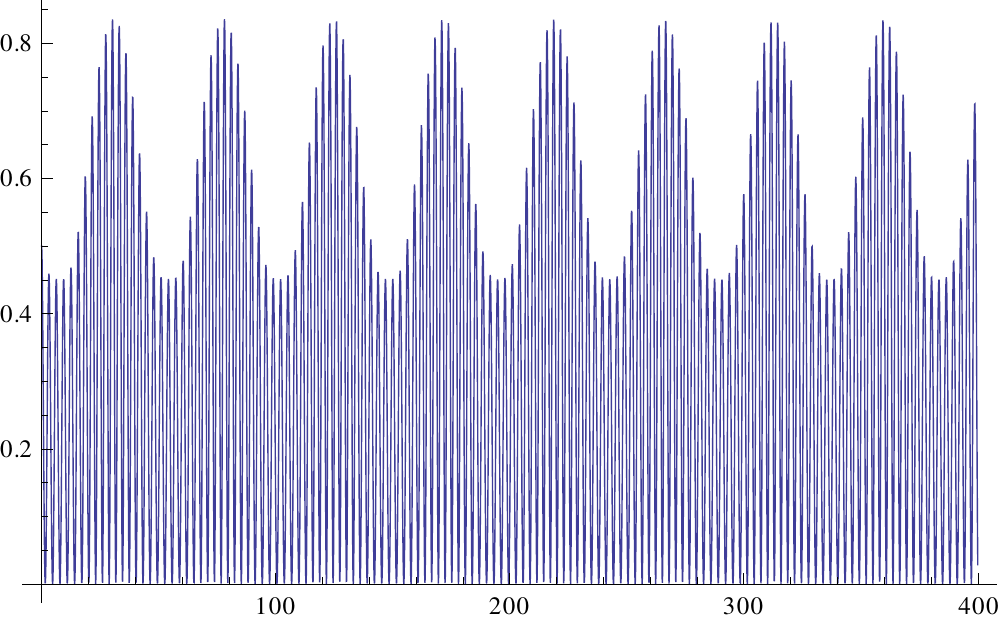}}
\put(25,0){\includegraphics[scale=0.4]{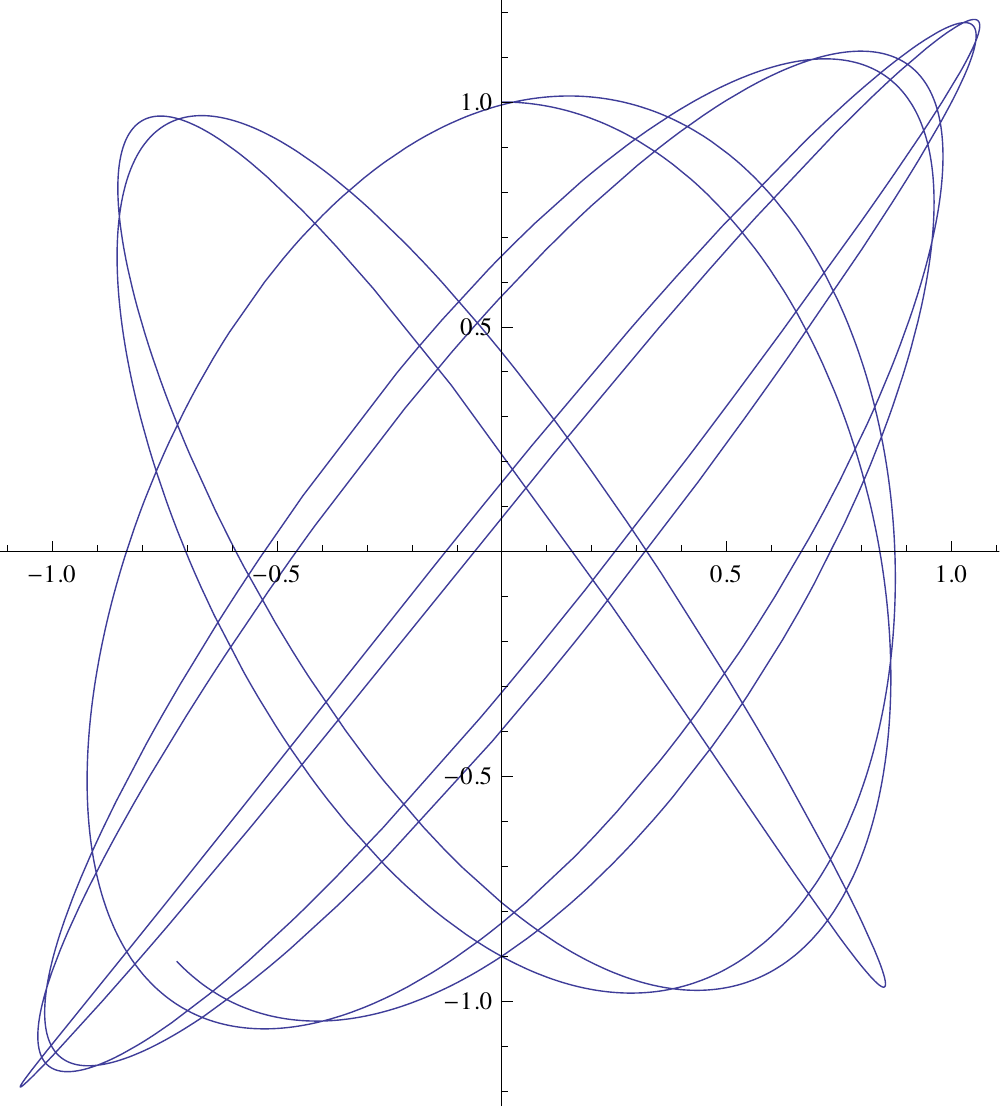}}
\put(90,0){\includegraphics[scale=0.4]{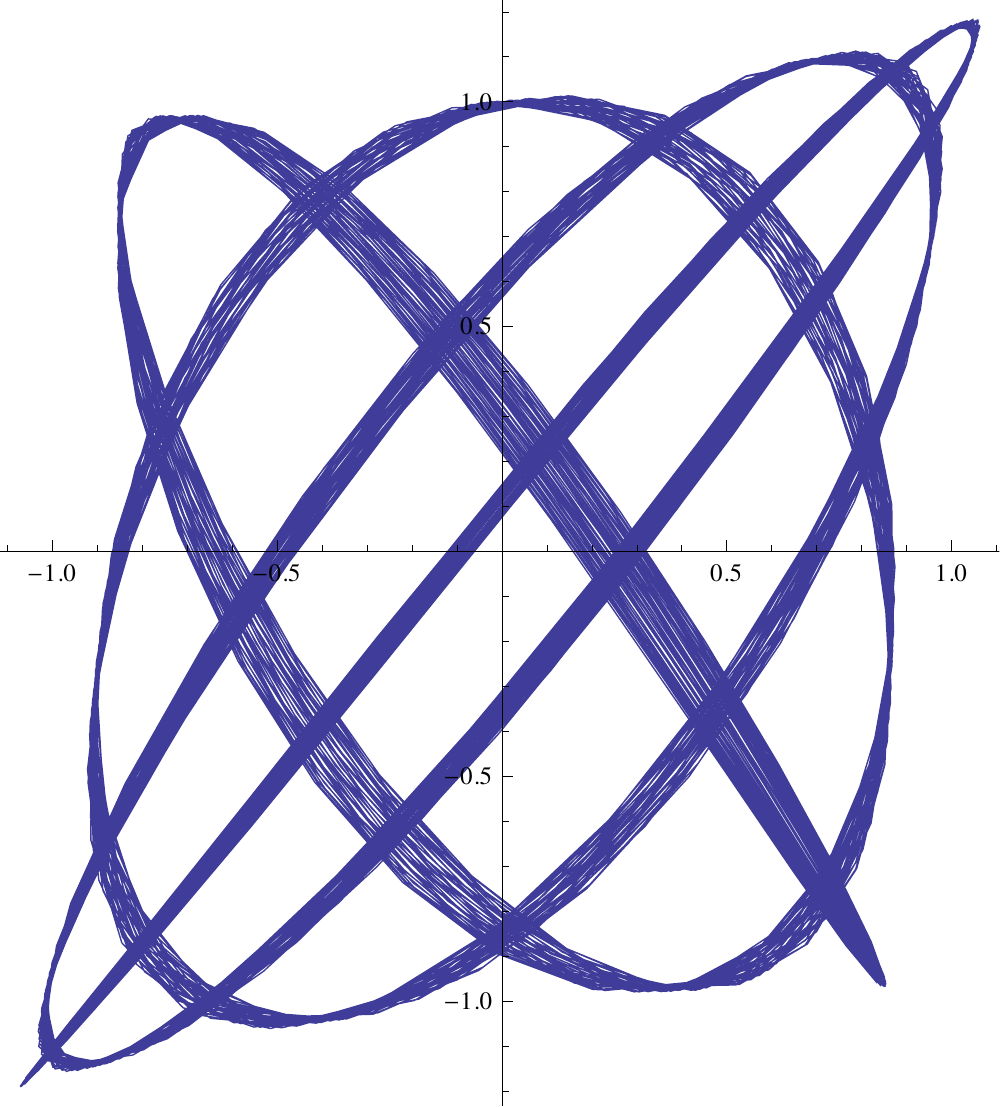}}
\put(150,0){\includegraphics[scale=0.4]{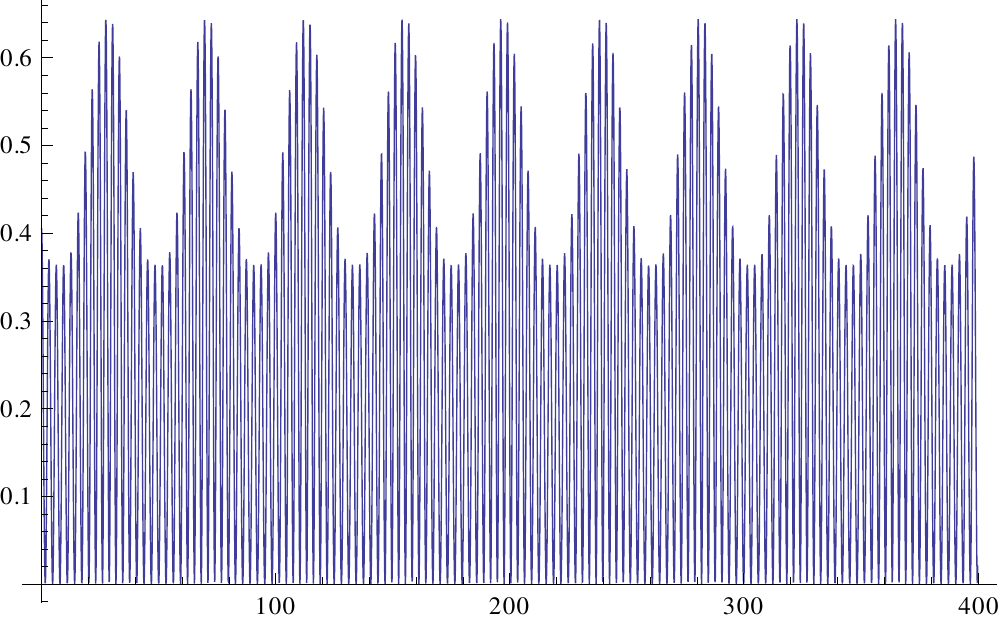}}

\put(73,83){$x'$}
\put(46,112){$y'$}

\put(139,83){$x'$}
\put(112,112){$y'$}

\put(74,24){$x'$}
\put(47,54){$y'$}

\put(140,24){$x'$}
\put(112,54){$y'$}

\put(145,92){$\frac{{\dot x}'^2}{2}$}
\put(201,59){$t$}

\put(145,30){$\frac{{\dot x}'^2}{2}$}
\put(201,-2){$t$}

\put(152,110){$^{\lambda=0.022}$}
\put(152,105){$^{x'(0)=0,~y'(0)=1}$}
\put(152,100){$^{{\dot x}'(0)=1,~{\dot y}'(0)=0}$}
\put(152,48){$^{\lambda=0.02299}$}
\put(152,43){$^{x'(0)=0,~y'(0)=1}$}
\put(152,38){$^{{\dot x}'(0)=0.9,~{\dot y}'(0)=0}$}

\end{picture}

\caption{\footnotesize Solutions of Eqs.\,(\ref{3.12})(\ref{3.13}) for
different values of the coupling constant $\lambda$ and different
initial conditions.
Left and middle: the trajectories in the $(x',y')$ space.
Right: The kinetic energy ${\dot x}'^2/2$ as function of time. The oscillations
within the envelope are so fine that they fill the diagram.}
\end{figure}
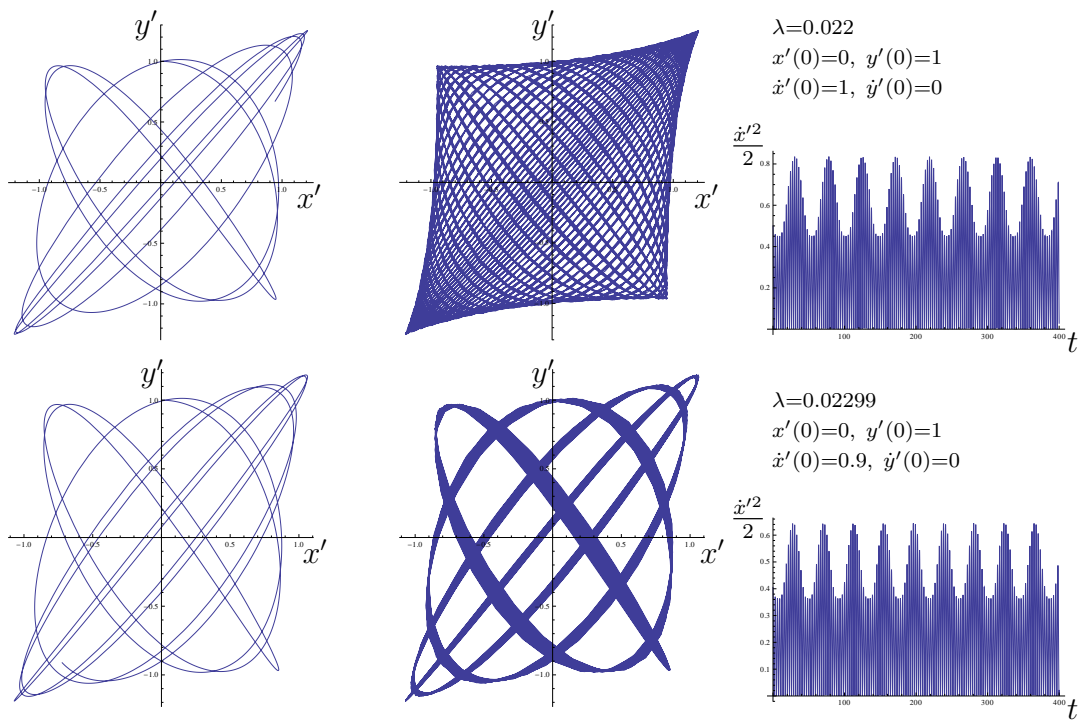

We see that the system is stable for sufficiently small coupling
constant $\lambda$ and  the initial velocity ${\dot x}'(0)$, ${\dot y}'(0)$.
If $\lambda$ is too high, the system is unstable (Fig.\,2, up).

\begin{figure}[h!]
\hs{3mm}
\begin{picture}(120,125)(25,0)

\put(25,61){\includegraphics[scale=0.4]{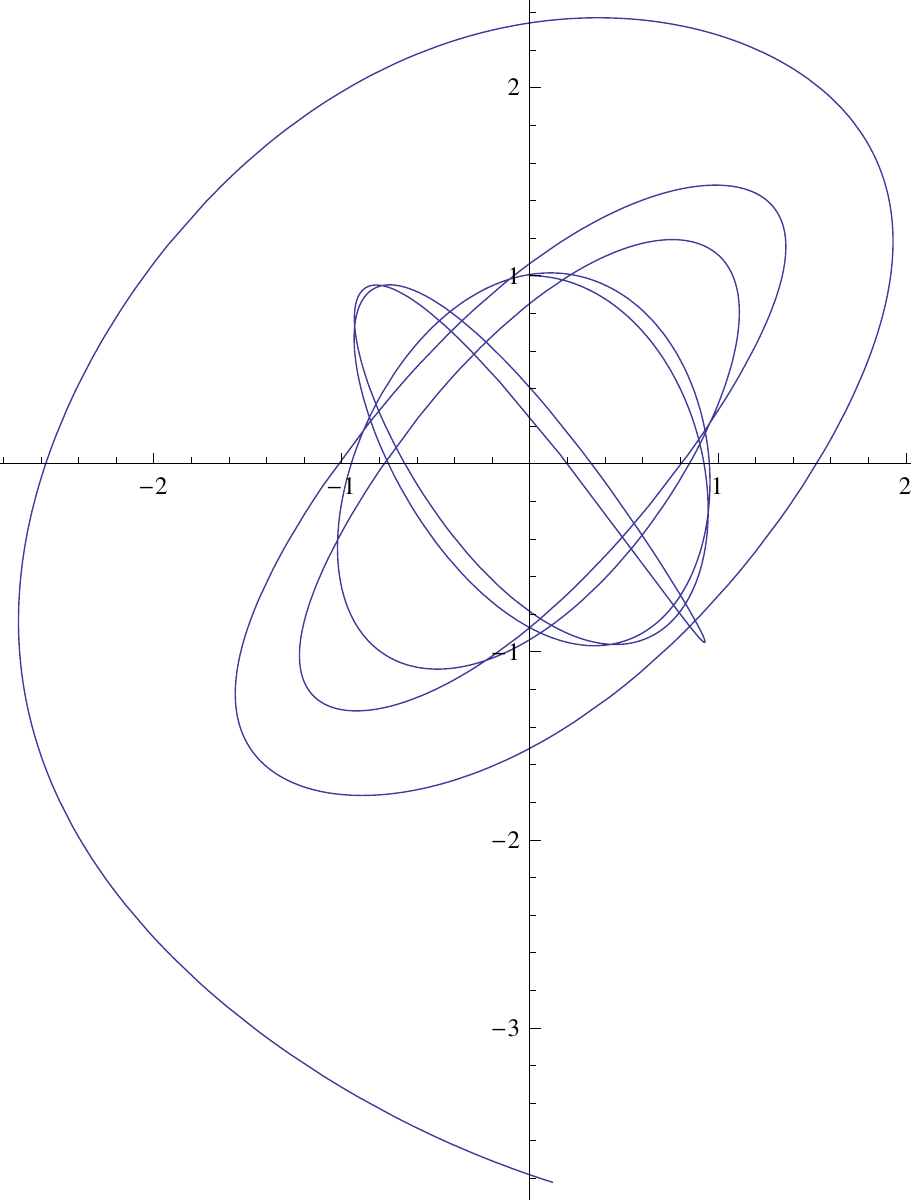}}
\put(90,61){\includegraphics[scale=0.4]{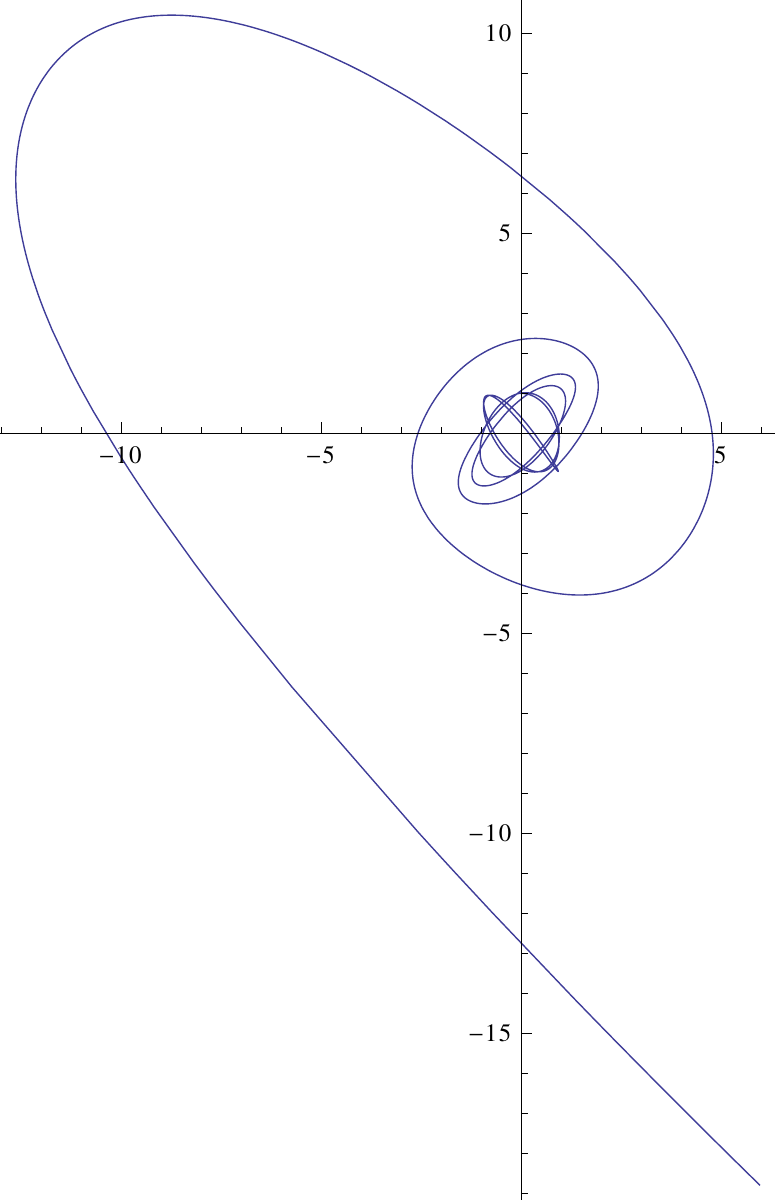}}
\put(150,61){\includegraphics[scale=0.4]{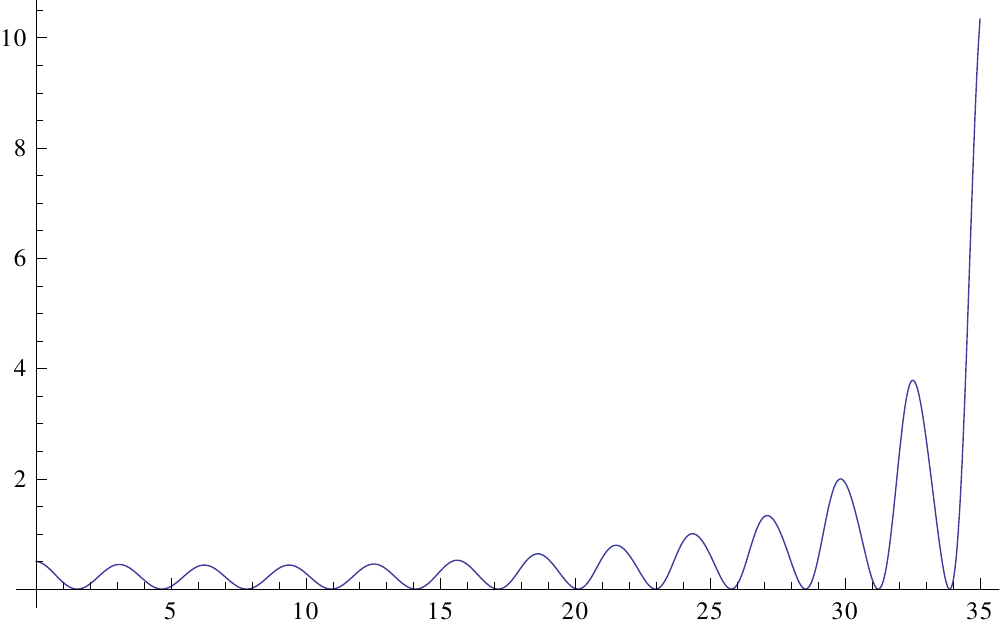}}
\put(25,0){\includegraphics[scale=0.4]{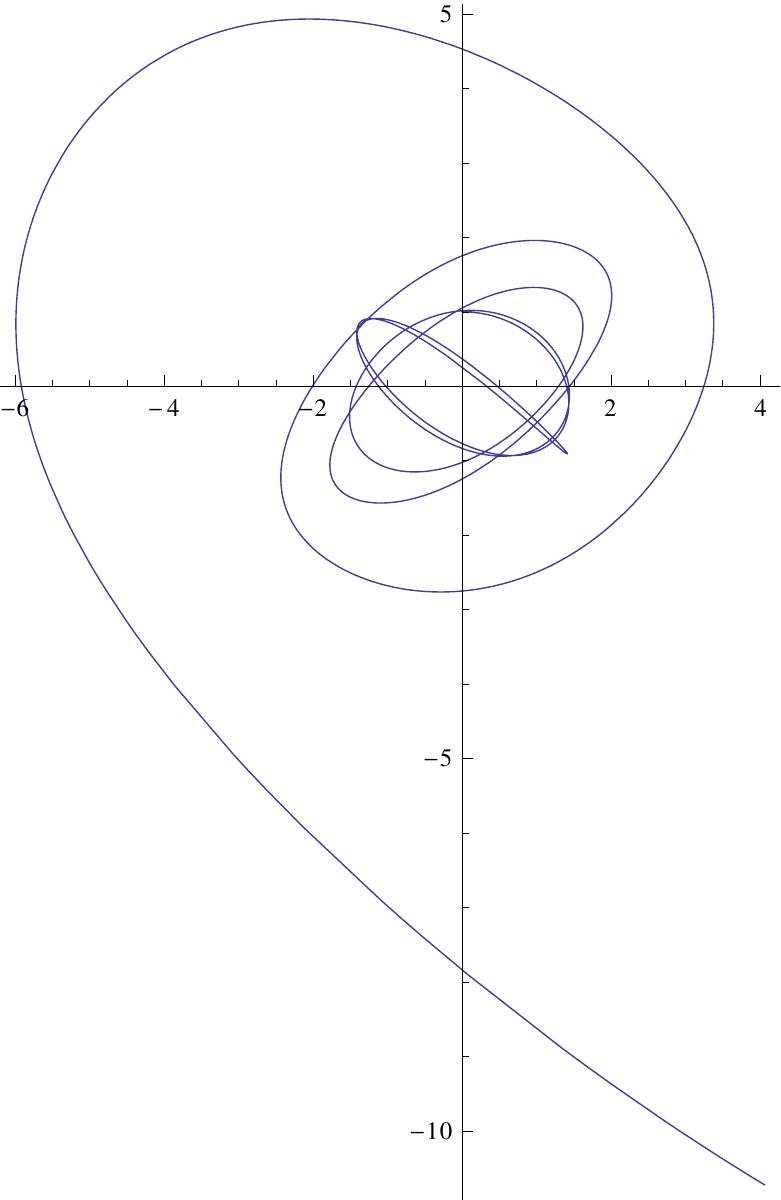}}
\put(90,0){\includegraphics[scale=0.4]{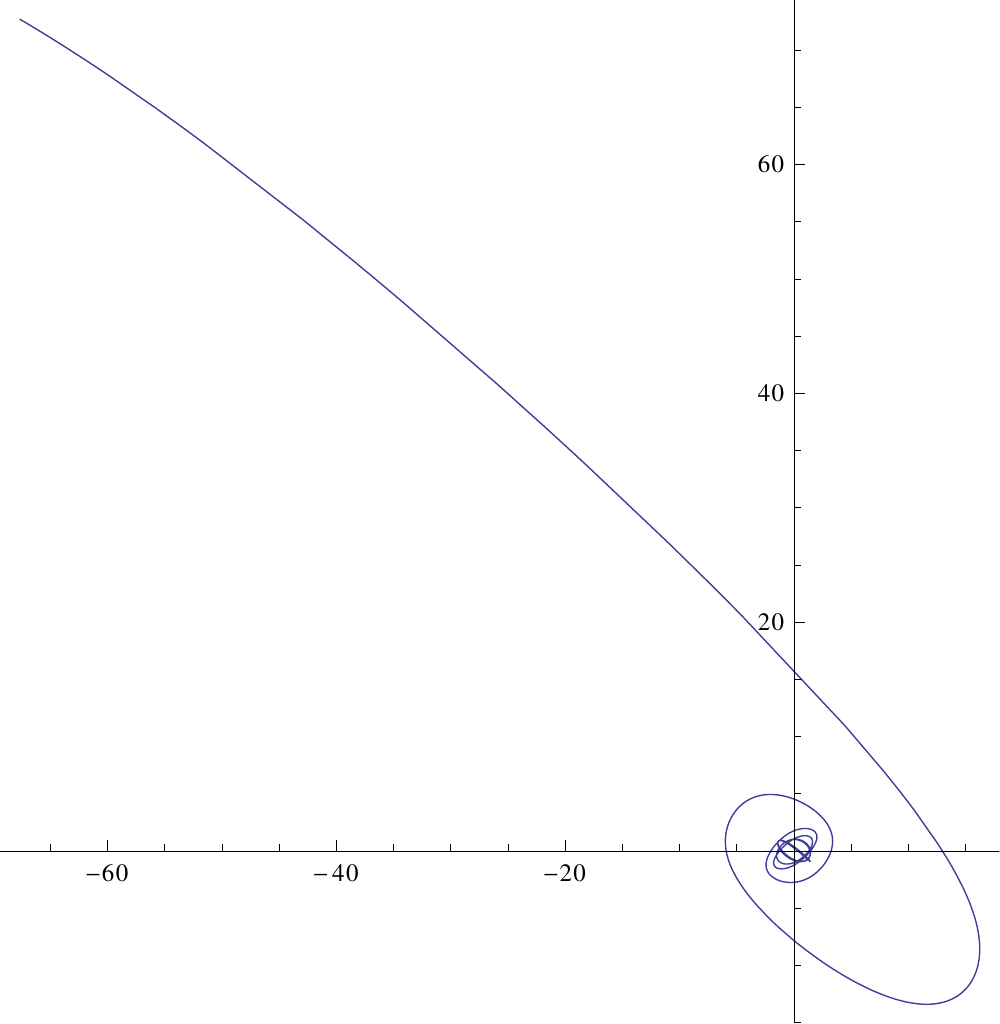}}
\put(150,0){\includegraphics[scale=0.4]{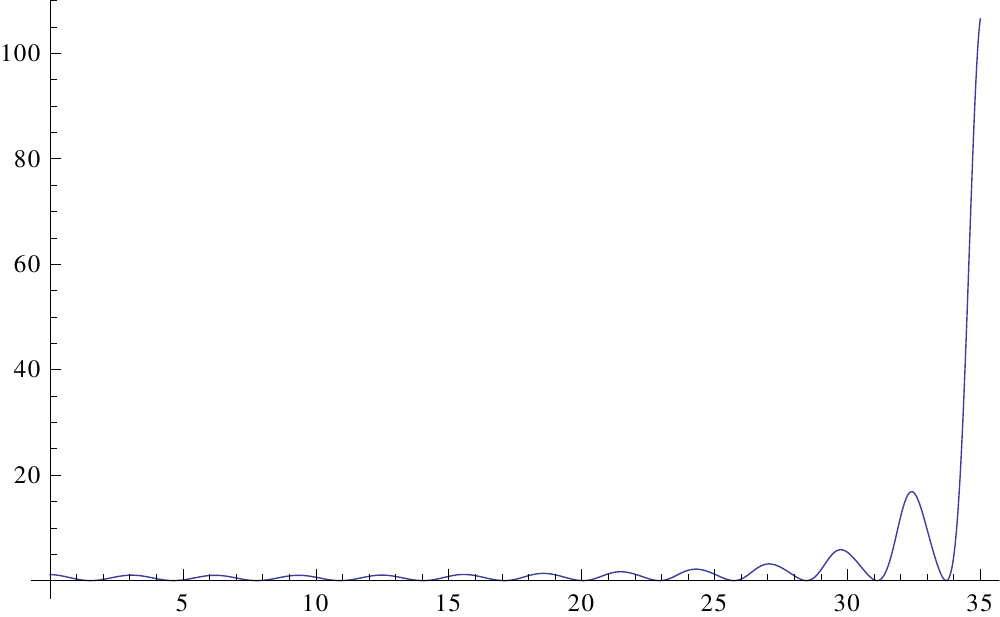}}

\put(73,95){$x'$}
\put(47,122){$y'$}

\put(130,96){$x'$}
\put(112,122){$y'$}

\put(65,37){$x'$}
\put(45,56){$y'$}

\put(140,10){$x'$}
\put(126,51){$y'$}

\put(145,92){$\frac{{\dot x}'^2}{2}$}
\put(201,59){$t$}

\put(145,30){$\frac{{\dot x}'^2}{2}$}
\put(201,-2){$t$}

\put(147,110){$^{\lambda=0.03}$}
\put(147,105){$^{x'(0)=0,~y'(0)=1}$}
\put(147,100){$^{{\dot x}'(0)=1,~{\dot y}'(0)=0}$}
\put(147,48){$^{\lambda=0.022}$}
\put(147,43){$^{x'(0)=0,~y'(0)=1}$}
\put(147,38){$^{{\dot x}'(0)=1.5,~{\dot y}'(0)=0}$}

\end{picture}

\caption{\footnotesize Up: By increasing the $\lambda$, the system becomes
unstable. The trajectory and the kinetic energy escape into infinity.
Down: Similarly, by increasing the initial velocity, the system also becomes
unstable. }
\end{figure}
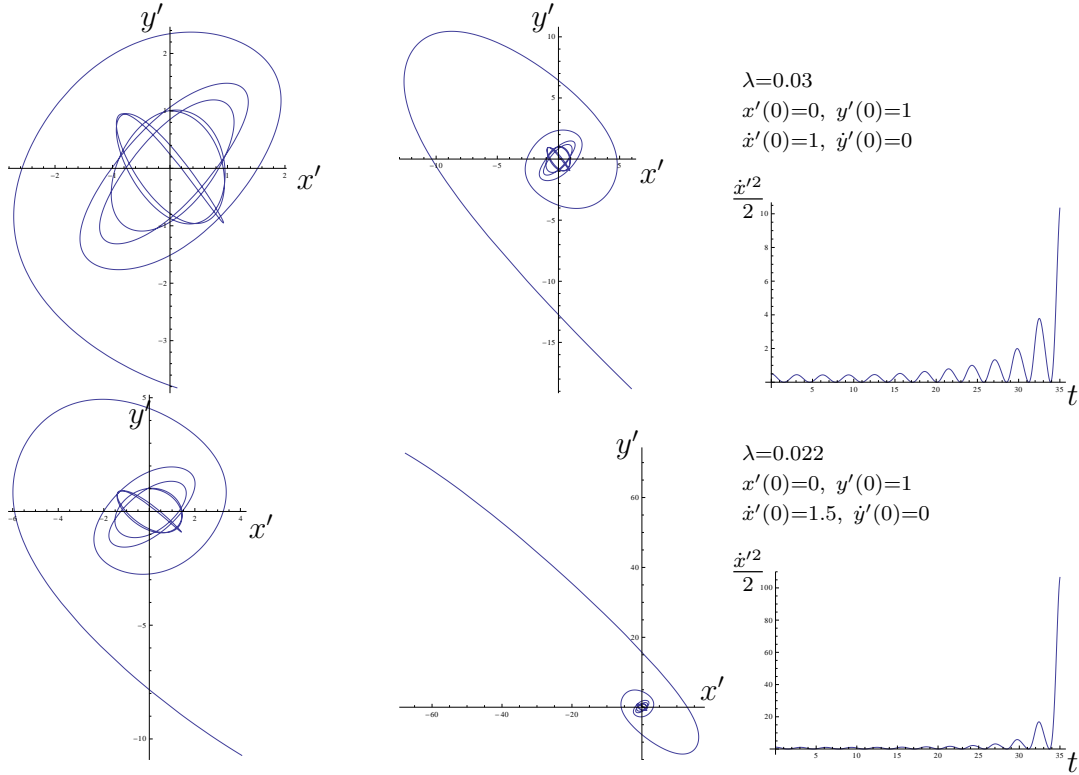

Similarly, the system is unstable at too high velocities (Fig.\,2, down). Close
to the critical value of $\lambda$, the system seems to be stable for
long time, but then it escapes into infinity (Fig.\,3).
A similar behaviour occurs close to the critical value of the initial
velocity.
\begin{figure}[h!]
\hs{3mm}
\begin{picture}(120,60)(25,0)

\put(25,0){\includegraphics[scale=0.4]{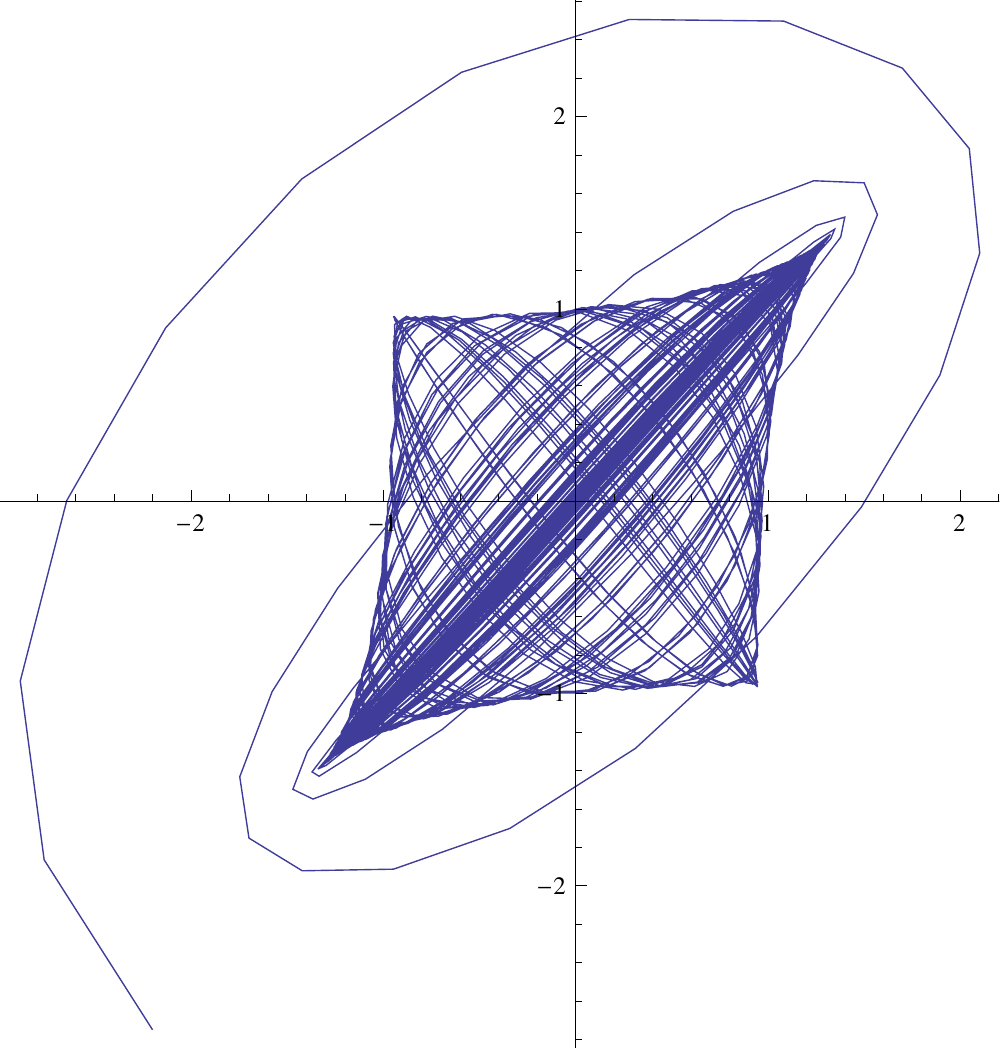}}
\put(90,0){\includegraphics[scale=0.4]{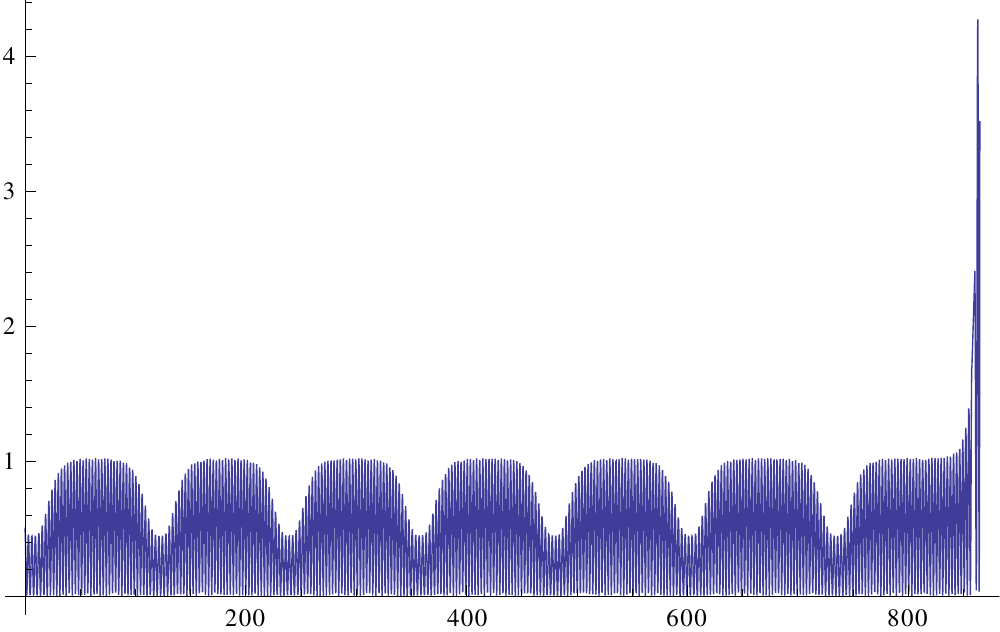}}
\put(150,0){\includegraphics[scale=0.4]{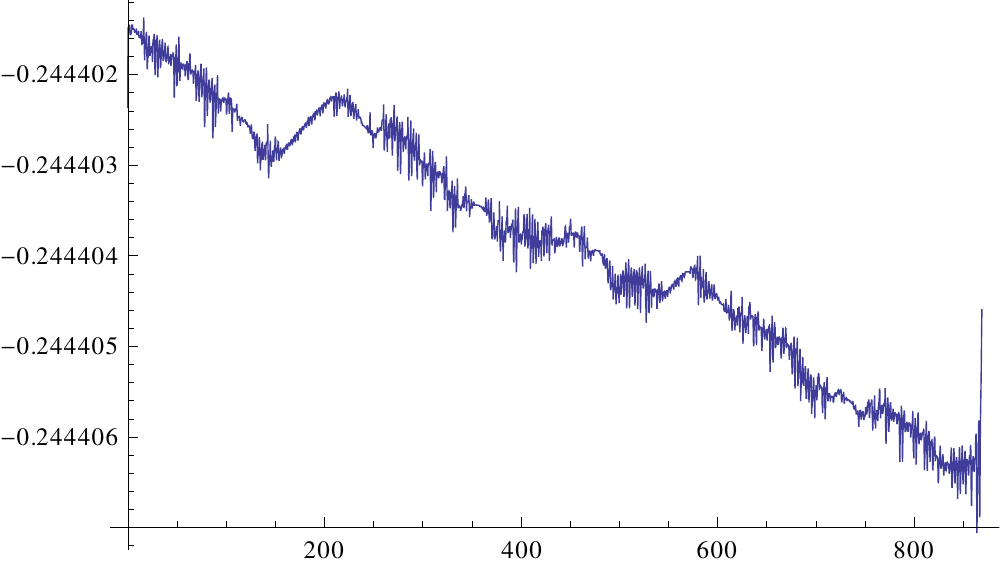}}

\put(74,23){$x$}
\put(51,53){$y$}

\put(85,30){$\frac{{\dot x}^2}{2}$}
\put(141,-2){$t$}

\put(152,30){$E_{\rm tot}$}
\put(201,-2){$t$}

\put(130,48){$^{\lambda=0.02299}$}
\put(130,43){$^{x'(0)=0,~y'(0)=1}$}
\put(130,38){$^{{\dot x}'(0)=0.999851,~{\dot y}'(0)=0}$}

\end{picture}

\caption{\footnotesize At certain values of $\lambda$ and the initial conditions,
the system behaves stably for a long time, before it finally escapes to infinity.
The total energy $E_{\text tot}$ remains constant within the numerical error.
}
\end{figure}
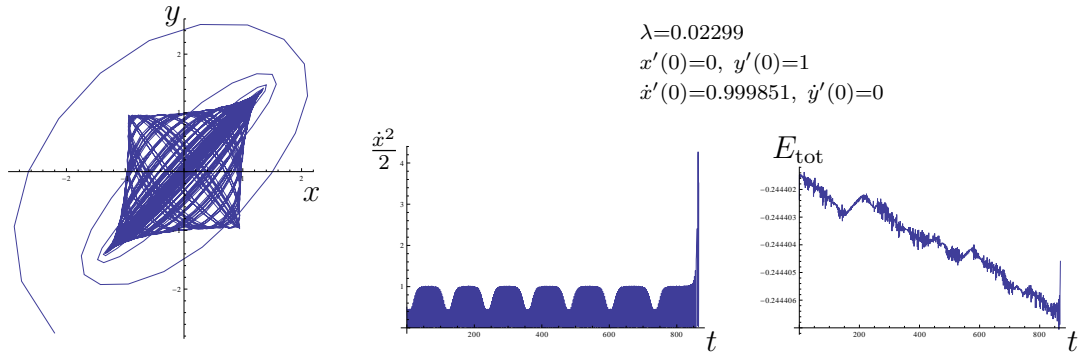

By just slightly decreasing the coupling constant
from that of figure 3, $\lambda =0.02299$, to $\lambda =0.0229$,
the system appears to be stable. We checked its stability up to $t= 2664$,
but we do not plot the solutions here, in order to not crowd the paper with
too many figures.

The interaction potential $\frac{\lambda}{4}(x'+y')^4$ runs into infinity.
More realistically, it should not run into infinity, but there should
be a cutoff. As a more realistic coupling term let us consider
$\frac{\lambda}{4}\text{sin}^4\, (x'+y')$, that leads to the
Lagrangian (\ref{3.19}) in which $u^4$ is replaced by $\text{sin}^4 u$.
 The equations of motion are then
\be
   {\ddot x'} +\omega_1^2 x' + \lambda \text{sin}^3\, (x'+y') 
   \text{cos} \,(x'+y')= 0,
\lbl{3.26}
\ee
\be
   {\ddot y'} +\omega_2^2 y'- \lambda \text{sin}^3\, (x'+y') 
   \text{cos} \,(x'+y')= 0
\lbl{3.27}
\ee
Such system is stable at all values of $\lambda>0$ and
initial velocity. We have checked this by performing many numerical runs.
In (Fig.\,4) we give two examples of numerical solutions. Later we will
demonstrate also analytically why the solutions of the system
(\ref{3.26}),(\ref{3.27}) are stable.

\setlength{\unitlength}{.8mm}

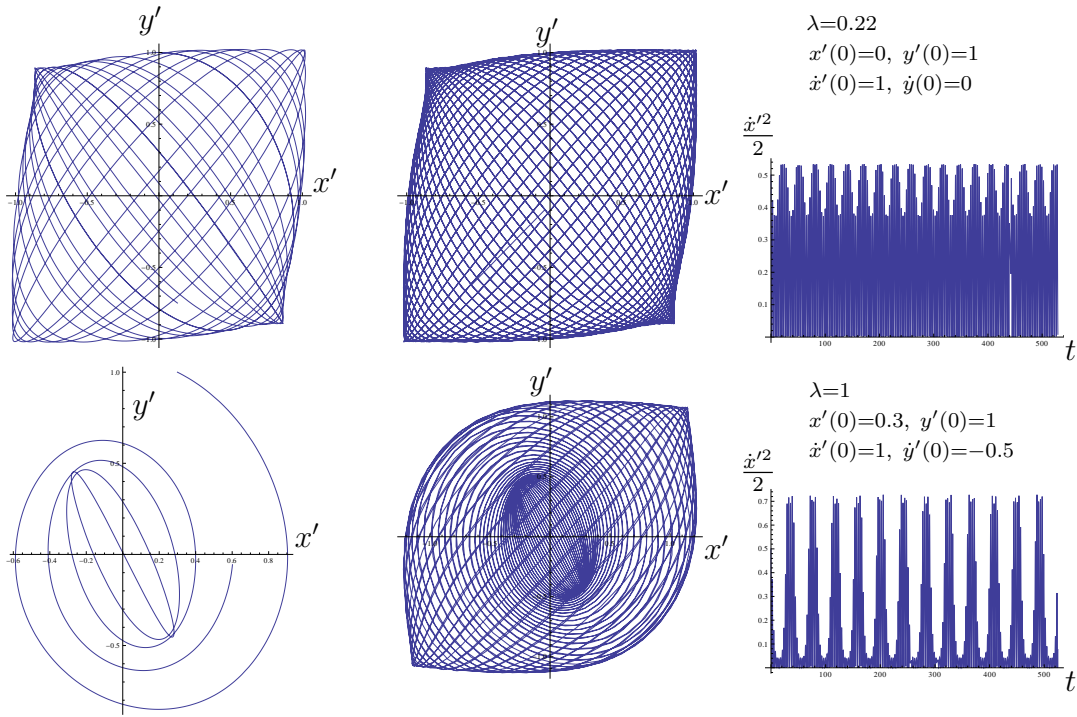
\begin{figure}[h!]
\hs{3mm}
\begin{picture}(120,115)(25,-5)

\put(25,55){\includegraphics[scale=0.4]{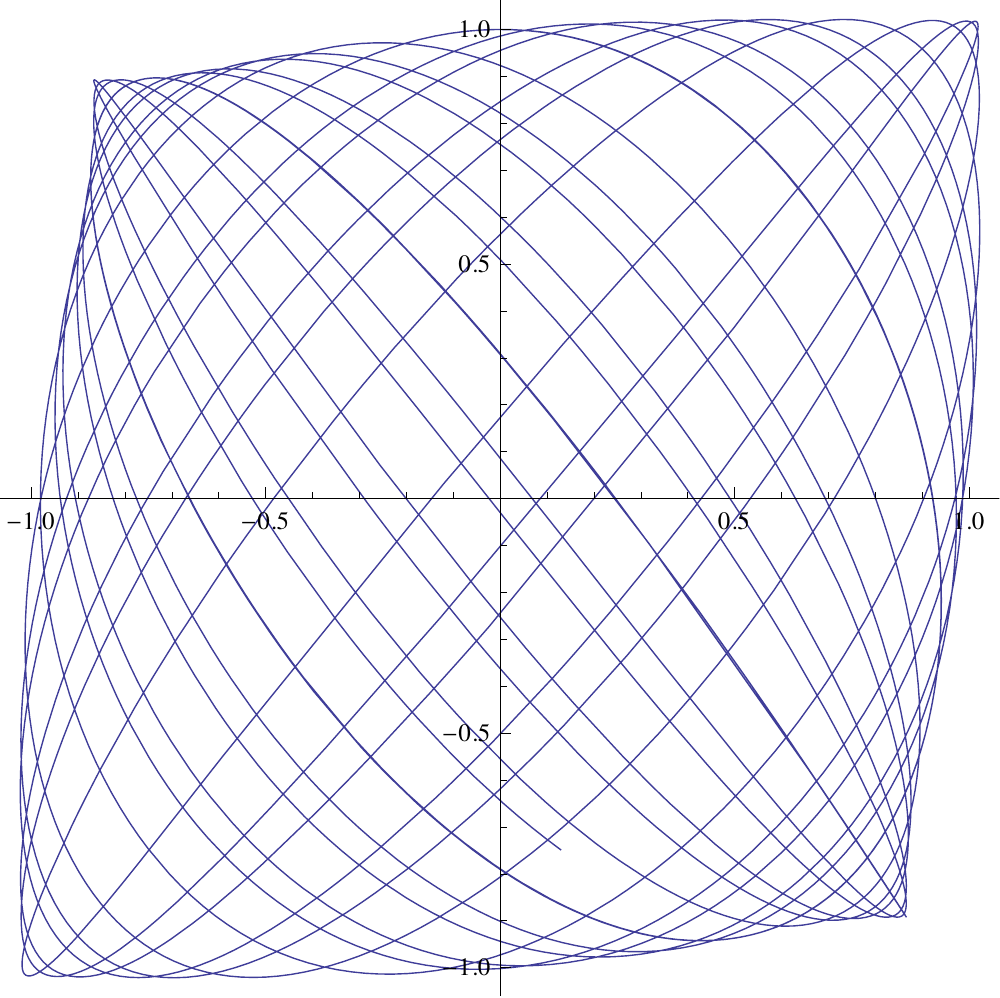}}
\put(90,55){\includegraphics[scale=0.4]{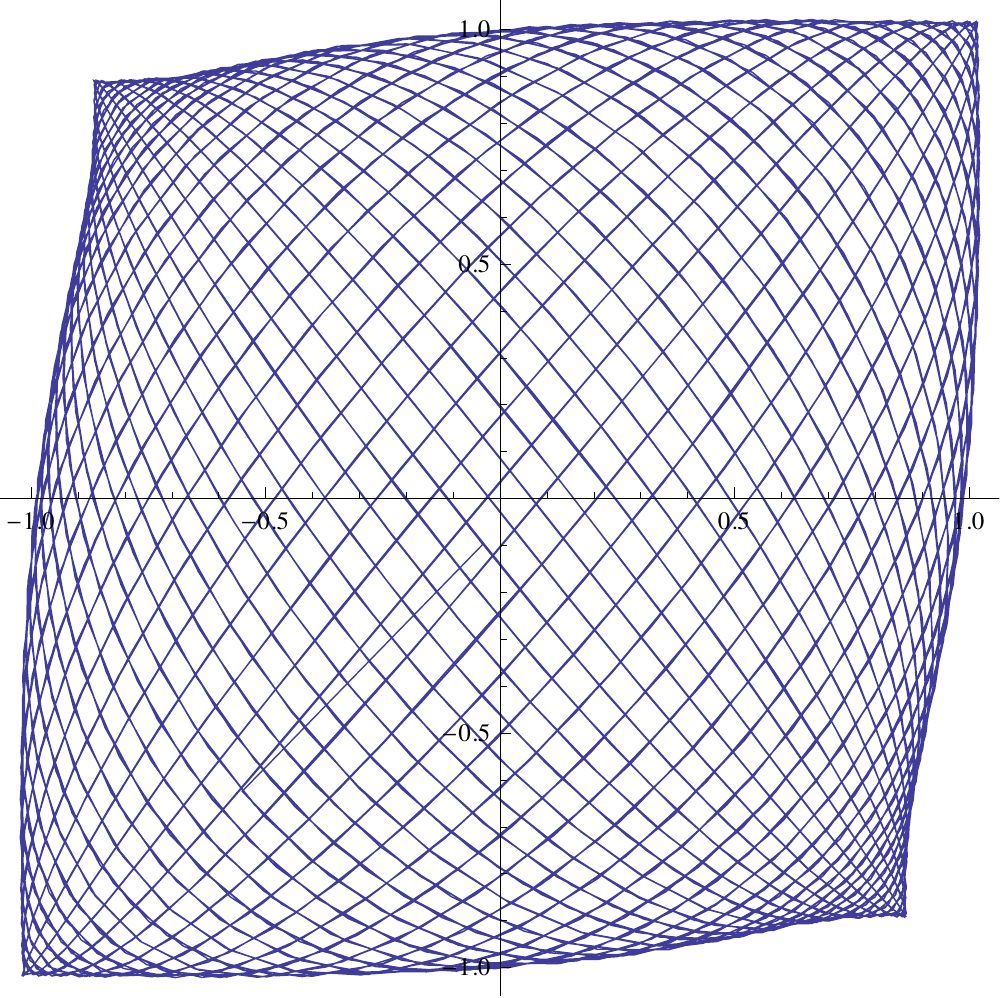}}
\put(150,55){\includegraphics[scale=0.4]{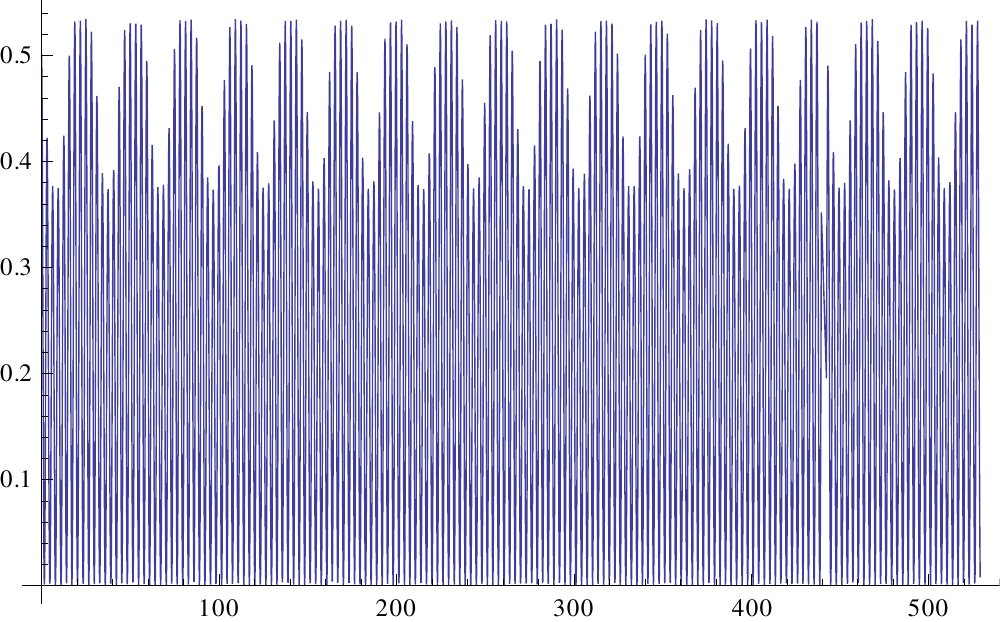}}
\put(25,-6){\includegraphics[scale=0.38]{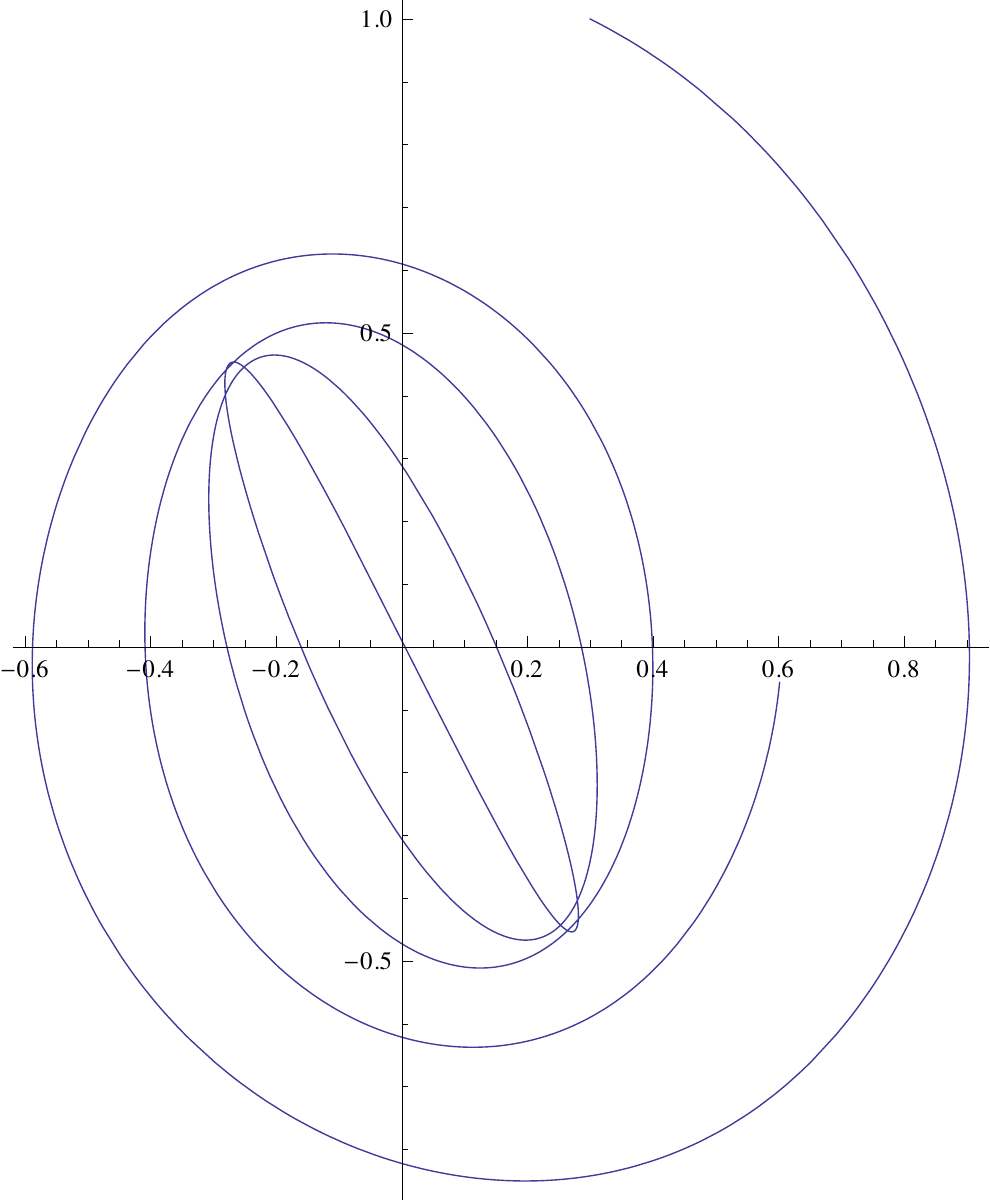}}
\put(90,0){\includegraphics[scale=0.4]{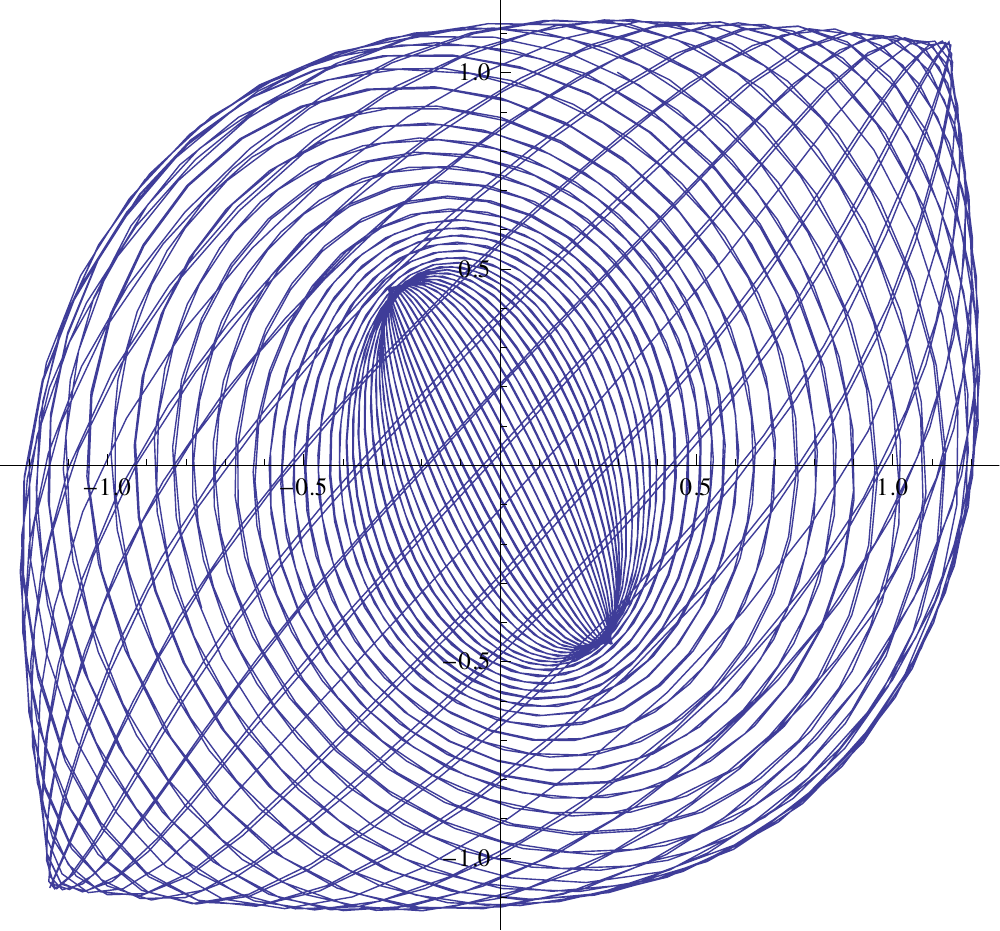}}
\put(150,0){\includegraphics[scale=0.4]{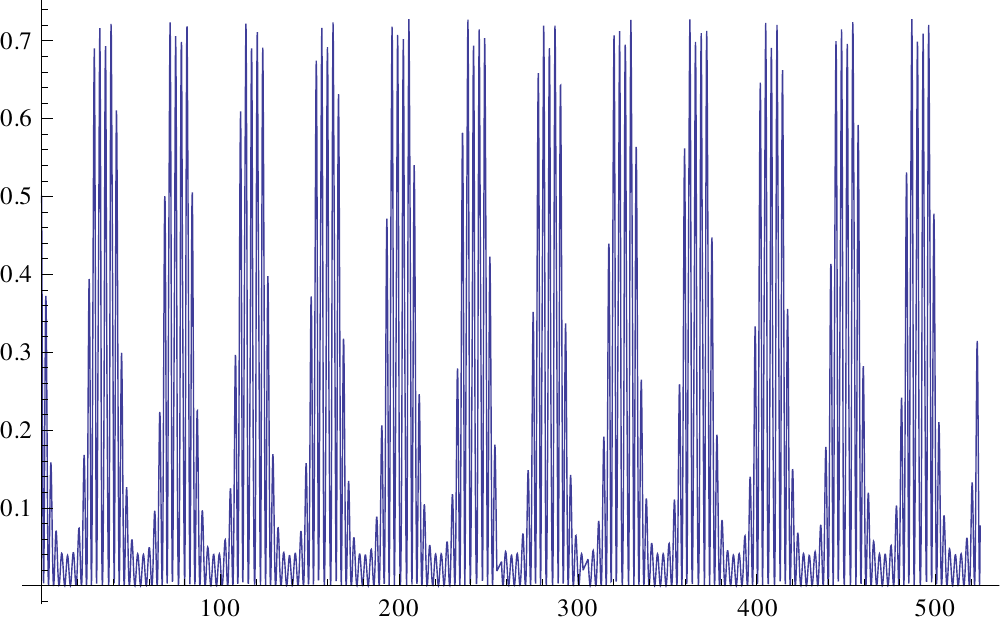}}

\put(76,81){$x'$}
\put(47,108){$y'$}

\put(141,79){$x'$}
\put(113,106){$y'$}

\put(73,22){$x'$}
\put(46,44){$y'$}

\put(141,20){$x'$}
\put(112,48){$y'$}

\put(147,89){$\frac{{\dot x}'^2}{2}$}
\put(201,53){$t$}

\put(147,33){$\frac{{\dot x}'^2}{2}$}
\put(201,-2){$t$}

\put(158,106){$^{\lambda=0.22}$}'
\put(157,101){$^{x'(0)=0,~y'(0)=1}$}
\put(157,96){$^{{\dot x}'(0)=1,~{\dot y}(0)=0}$}
\put(157,45){$^{\lambda=1}$}
\put(157,40){$^{x'(0)=0.3,~y'(0)=1}$}
\put(157,35){$^{{\dot x}'(0)=1,~{\dot y}'(0)=-0.5}$}

\end{picture}

\caption{\footnotesize Solutions to the equations of motion (\ref{3.26}),
(\ref{3.27}), in which the quartic interaction $\frac{\lambda}{4} (x'+y')^4$
is replaced by $\frac{\lambda}{4}\text{sin}^4\, (x'+y')$. The system
is now stable for all positive values of $\lambda$.}
\end{figure} 

Another possible generalization is in replacing the Lagrangian (\ref{3.11}) with
\be
    L=\mbox{$\frac{1}{2}$}(m_1 {\dot x}'^2 - m_2 {\dot y}'^2) - \mbox{$\frac{1}{2}$}
    (\omega_1^2 x'^2 - \omega_2^2 y'^2) - \frac{\lambda}{4} (x'+y')^4
 \lbl{3.28}
\ee
where $m_1$ and $m_2$ are now two different ``masses". In terms of the
variables $u$, $v$, we have
\be
    L = \mbox{$\frac{1}{2}$} \left [ m ({\dot u}^2 + {\dot v}^2) + 2 M 
    {\dot u} {\dot v} + \rho_1 (u^2 + v^2) - 2 \mu_1 u v \right ]
    - \lambda u^4 ,
\lbl{3.29}
\ee
where
\be
     m= \mbox{$\frac{1}{2}$} (m_1 - m_2) , ~~~~ 
     M = \mbox{$\frac{1}{2}$} (m_1 + m_2)
\lbl{3.30}
\ee
The equations of motion are now
\be
    m {\ddot u} + M {\ddot v} - \rho_1 u + \mu_1 v + 4 \lambda u^3 = 0
\lbl{3.31}
\ee
\be
    m {\ddot v} + M {\ddot u} - \rho_1 v + \mu_1 u =0 .
\lbl{3.32}
\ee
The corresponding 4th order equation is
\be
u^{(4)} M (M^2-m^2) + 2 {\ddot u} M (\mu_1 M + \rho_1 m) 
+ u M (\mu_1^2 - \rho_1^2)
      + 4 M \rho_1\lambda u^3 - 4 M m \lambda \frac{\dd^2}{\dd t^2}
     \left ( u^3 \right ) = 0
\lbl{3.33}
\ee
This is a deformed version of the equation (\ref{3.17}) for the interacting
PU oscillator. By taking $m=0$, $M=1$, we obtain the ordinary PU oscillator of
Eq.\,(\ref{3.17}). 

Examples of numerical solutions to the equations of motion 
\be
    m_1 {\ddot x}' + \omega_1^2 x' + \lambda (x'+y')^3 = 0,
\lbl{3.34}
\ee
\be
     m_2 {\ddot y} + \omega_2^2 y' - \lambda (x'+y')^3 = 0,
\lbl{3.35}
\ee
derived from the Lagrangian (\ref{3.28}), are given in Fig.\,5.
Whilst in the case of equal masses, $m_1=m_2=1$, the system
is unstable at $\lambda=0.03$ and higher, we see that for different masses,
$m_1<m_2$, the system is stable regardless of the values of $\lambda>0$
and initial velocities. This has been confirmed in many numerical runs
that we have done. Only a small sample is shown in Fig.\,5.
That for unequal masses the system becomes stable we previously observed
in Ref.\,\ci{PavsicUltrahyper}, where we studied an analogous system of two
oscillators, but with a different coupling term, namely,
$\frac{\lambda}{4}(x^2-y^2)^2$, which is a special case of that
considered in Ref.\,\ci{Ilhan}. However, such a coupling term does not
correspond to a quartic self-interaction of the PU oscillator, because the
coupling term $\lambda u^4$ in (\ref{3.14}) is then replaced by $\lambda u^2 v^2$,
which does not lead to Eq.\,(\ref{3.18}), but to a more complicated equation
with non-linear terms.

\setlength{\unitlength}{.8mm}

\begin{figure}[h!]
\hs{3mm}
\begin{picture}(120,115)(25,0)

\put(25,61){\includegraphics[scale=0.4]{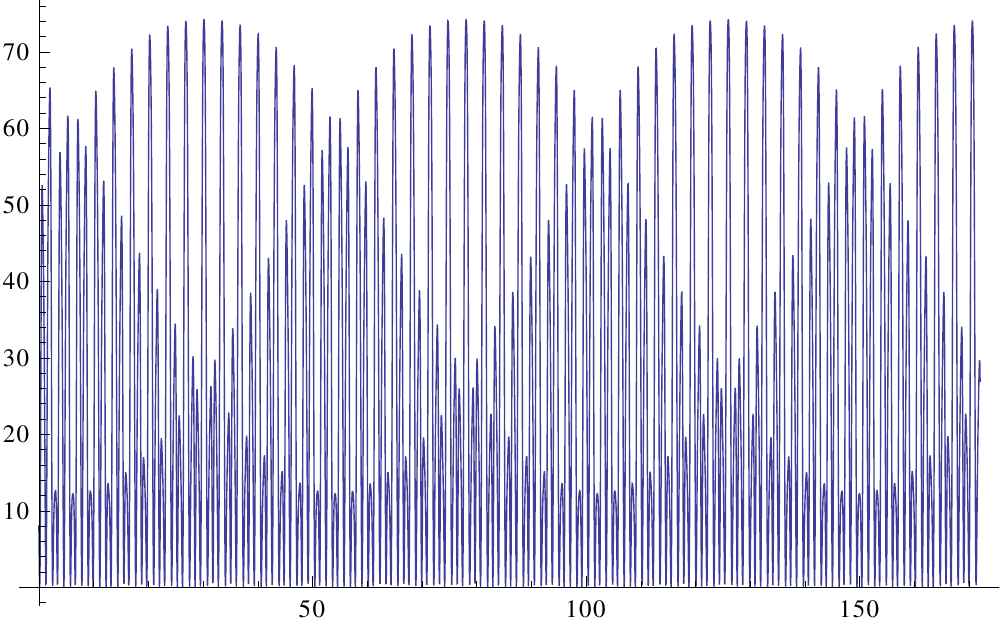}}
\put(90,61){\includegraphics[scale=0.4]{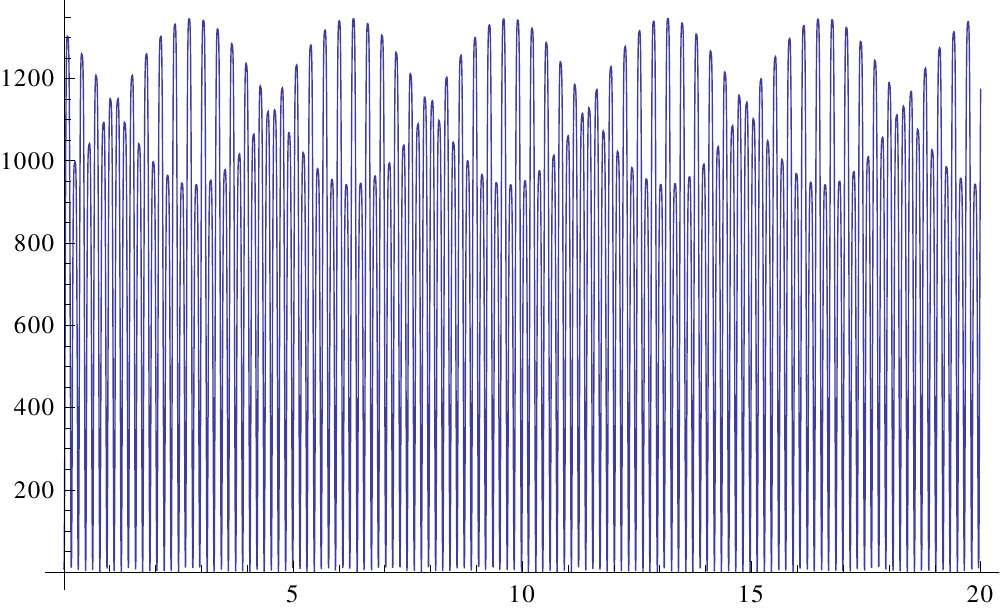}}
\put(150,61){\includegraphics[scale=0.4]{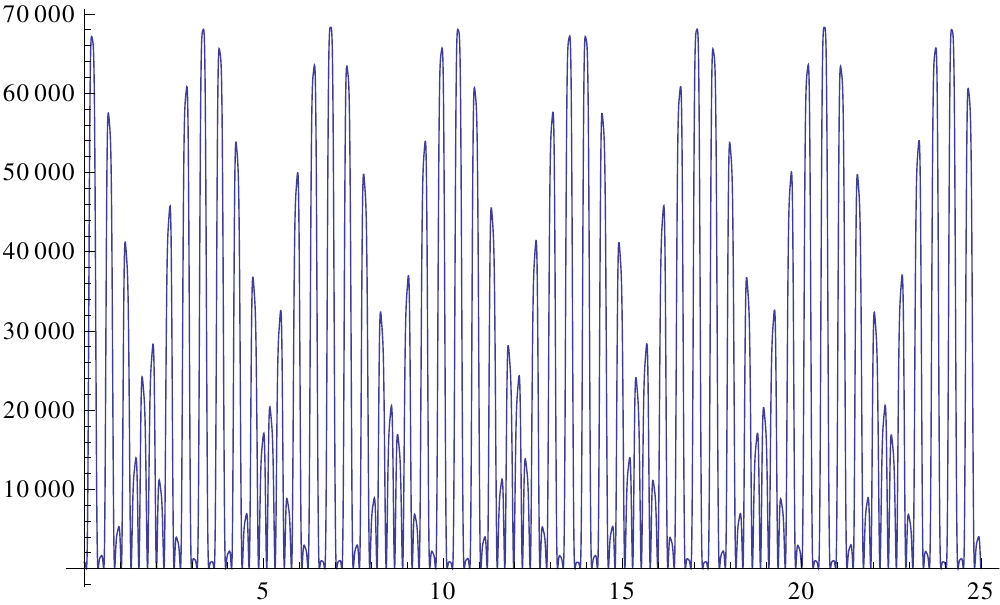}}
\put(25,0){\includegraphics[scale=0.4]{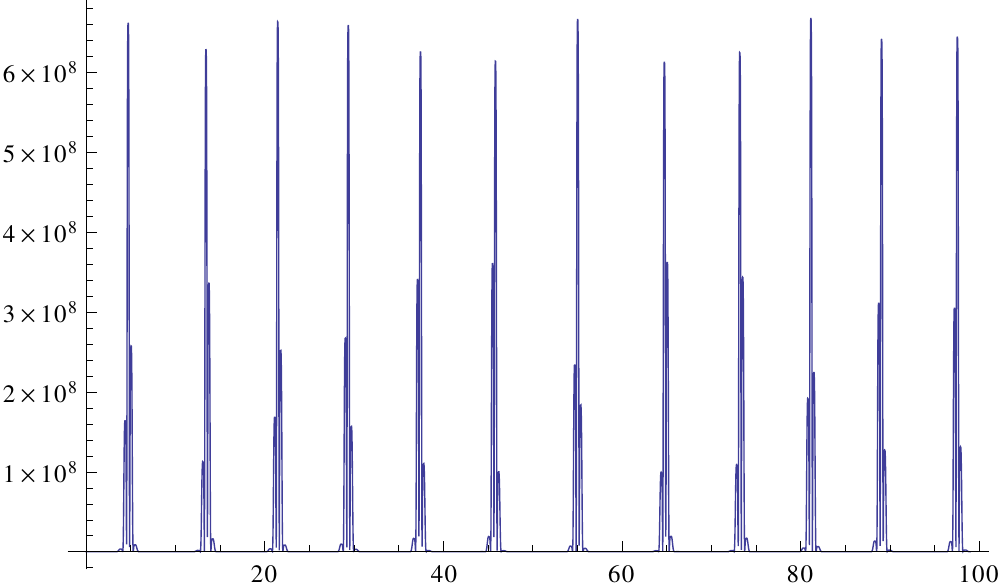}}
\put(90,0){\includegraphics[scale=0.4]{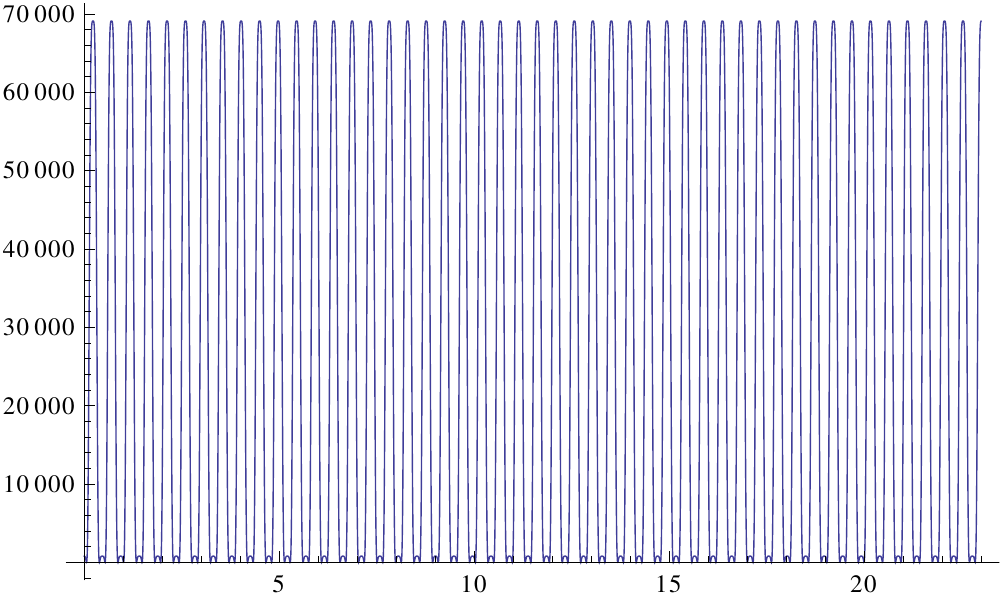}}
\put(150,0){\includegraphics[scale=0.4]{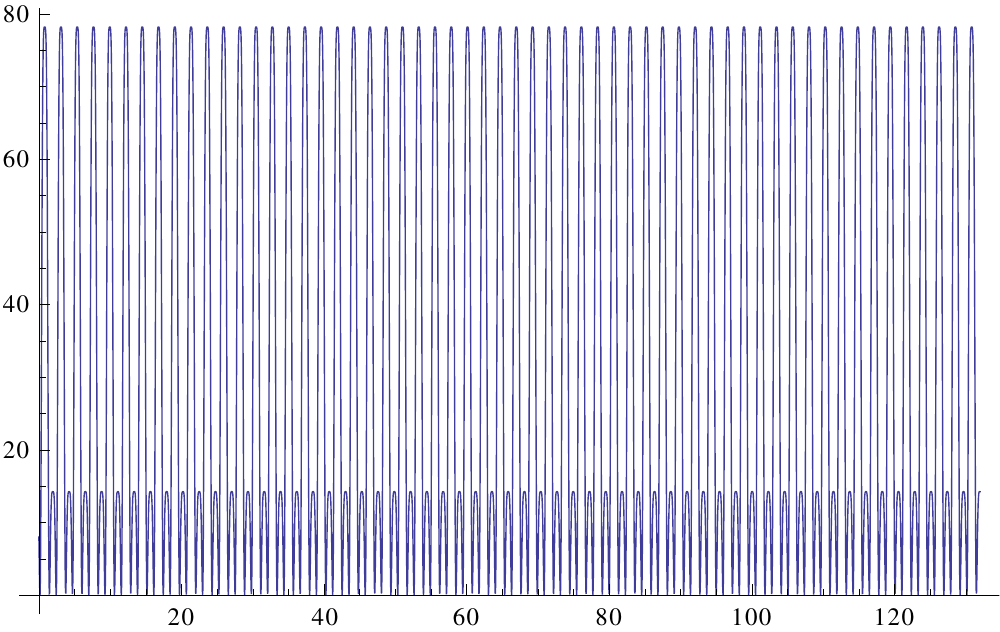}}

\put(21,92){$\frac{{\dot x}'^2}{2}$}
\put(76,59){$t$}

\put(87,92){$\frac{{\dot x}'^2}{2}$}
\put(141,59){$t$}

\put(145,92){$\frac{{\dot x}'^2}{2}$}
\put(201,59){$t$}

\put(21,30){$\frac{{\dot x}'^2}{2}$}
\put(76,-2){$t$}

\put(85,30){$\frac{{\dot x}'^2}{2}$}
\put(142,-2){$t$}

\put(145,30){$\frac{{\dot x}'^2}{2}$}
\put(202,-2){$t$}

\put(25,110){$^{\lambda=5,~m_1=0.7,~m_2=1.3}$}
\put(25,105){$^{x'(0)=0.3,~y'(0)=1}$}
\put(25,100){$^{{\dot x}'(0)=4,~{\dot y}'(0)=-0.5}$}
\put(30,95){$^{\omega_1=1,~\omega_2=\sqrt{1.5}}$}

\put(91,110){$^{\lambda=500,~m_1=0.7,~m_2=1.3}$}
\put(91,105){$^{x'(0)=0.3,~y'(0)=1}$}
\put(91,100){$^{{\dot x}'(0)=4,~{\dot y}'(0)=-0.5}$}
\put(96,95){$^{\omega_1=1,~\omega_2=\sqrt{1.5}}$}

\put(152,110){$^{\lambda=5,~m_1=0.7,~m_2=1.3}$}'
\put(152,105){$^{x'(0)=0.3,~y'(0)=1}$}
\put(152,100){$^{{\dot x}'(0)=40,~{\dot y}'(0)=55}$}
\put(157,95){$^{\omega_1=1,~\omega_2=\sqrt{1.5}}$}
\put(25,48){$^{\lambda=5,~m_1=0.99,~m_2=1.01}$}
\put(25,43){$^{x'(0)=0.3,~y'(0)=1}$}
\put(25,38){$^{{\dot x}'(0)=1,~{\dot y}'(0)=-0.5}$}
\put(30,33){$^{\omega_1=1,~\omega_2=\sqrt{1.5}}$}

\put(91,48){$^{\lambda=5,~m_1=0.7,~m_2=1.3}$}
\put(91,43){$^{x'(0)=0.3,~y'(0)=1}$}
\put(91,38){$^{{\dot x}'(0)=40,~{\dot y}'(0)=55}$}
\put(96,33){$^{\omega_1=\omega_2=0}$}

\put(152,48){$^{\lambda=5,~m_1=0.7,~m_2=1.3}$}
\put(152,43){$^{x'(0)=0.3,~y'(0)=1}$}
\put(152,38){$^{{\dot x}'(0)=4,~{\dot y}'(0)=-0.5}$}
\put(157,33){$^{\omega_1=\omega_2=0}$}

\end{picture}

\caption{\footnotesize Solutions of Eqs.\,(\ref{3.34})(\ref{3.35}) for
different values of the coupling constant $\lambda$ and different
initial conditions. We show here the kinetic energy ${\dot x}'^2/2$ as
function of time.}
\end{figure}
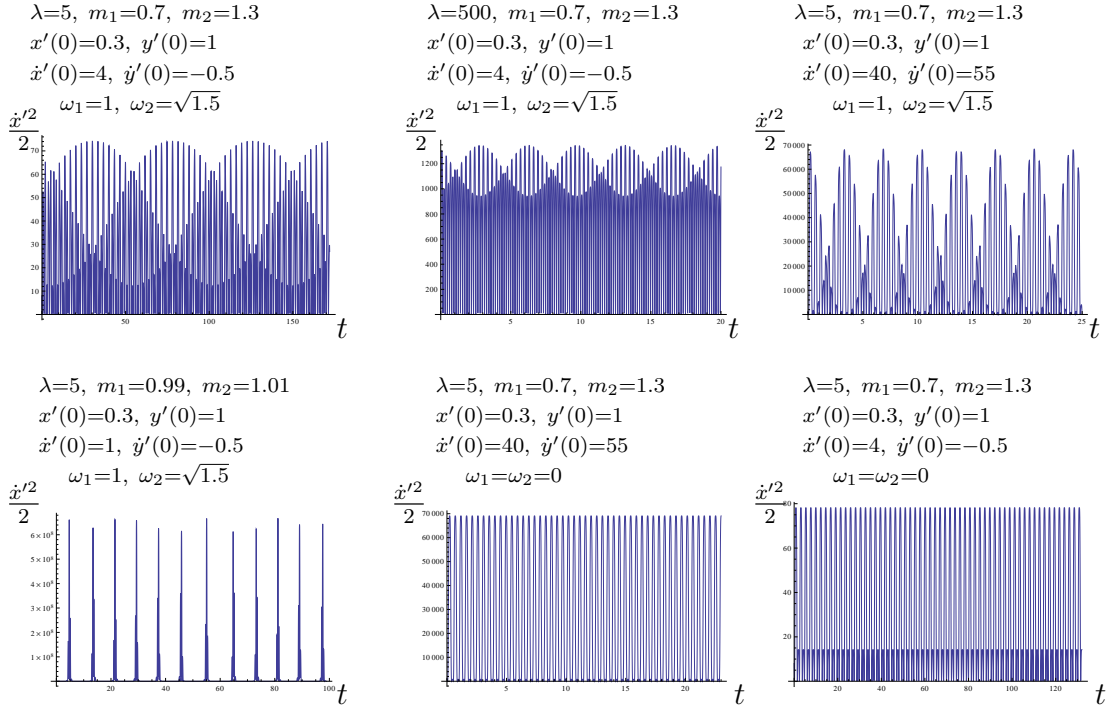

To see why the system with different masses is stable, let us inspect the
equations of motion (\ref{3.34}),(\ref{3.35}). We already know that at small
values of $\lambda$, the system is stable. At large values of $\lambda$,
we can neglect the terms $\omega_1^2 x'$ and $\omega_2^2 y'$. Equations
of motion are then
\be
   {\ddot x}' + \frac{1}{m_1} \lambda (x'+y')^3 = 0,
\lbl{3.36}
\ee
\be
{\ddot y}' - \frac{1}{m_2} \lambda (x'+y')^3 = 0.
\lbl{3.37}
\ee
Taking the sum and the difference of the latter equations, we obtain
\be
   {\ddot \xi} + \left (\frac{1}{m_1} - \frac{1}{m_2} \right ) \lambda \xi^3=0,
\lbl{3.38}
\ee
\be
   {\ddot \eta} + \left (\frac{1}{m_1} + \frac{1}{m_2} \right ) \lambda \xi^3=0,
\lbl{3.39}
\ee
where $\xi=x'+y'$ and $\eta=x'-y'$.

{\it In the case of unequal masses}, $m_1 < m_2$, $\lambda >0$,
Eq.\,(\ref{3.38}) describes the quartic oscillator with the potential
$\frac{\lambda}{4} \xi^4 (1/m_1-1/m_2)$, which has stable, oscillatory solutions. Then,
Eq,\,(\ref{3.39}) also has stable, oscillatory, solutions. Stability is
maintained in the presence of the terms $\omega_1^2 x'$ and $\omega_2^2 y'$.

{\it In the case of equal masses}, $m_1=m_2$, Eq.\,(\ref{3.38}) becomes
${\ddot \xi}=0$, with the solution $\xi=\xi_0 + c_1 t$. Then the general solution of
(\ref{3.39}) is a runaway function
\be
   \eta = - \frac{2}{m_1} \frac{\lambda}{20} (\xi_0 + c_1 t)^5 + c_2 t .
\lbl{3.40}
\ee
In the presence of the terms $\omega_1^2 x$ and $\omega_2^2 y$, the above
runaway behavior is modulated by oscillations.

Solutions to the system described by the Lagrangian (\ref{3.28}) are
stable, if $m_1 < m_2$, $\lambda >0$. If $m_1=m_2$, then the solutions are
stable at sufficiently small $\lambda$, whereas at higher values of
$\lambda$, they are unstable. (Fig.\,2).

If instead of $\frac{1}{4}(x'+y')$ we take the interaction term
$\frac{1}{4} \text{sin}^4\,(x'+y')$, we have stability even in the
case $m_1=m_2$. The equations of motion are then
\be
     {\ddot \xi}=0~,~~~~~~{\ddot \eta} + \frac{2}{m_1} \lambda
     \,\text{sin}^3 \, \xi \, \text{cos}\, \xi =0,
\lbl{3.41}
\ee
the general solution being
\be  
    \xi=\xi_0 + c_1 t~,~~~{\dot \eta}= -\frac{2}{m_1} \frac{1}{4 c_1} \lambda
    \,\text{sin}^4 \,(\xi_0 + c_1 t)~,
\lbl{3.42}
\ee
\be
    \eta = -\frac{2}{m_1} \frac{1}{128 c_1^2} \lambda
    \left [12 (\xi_0 + c_1 t) -8\, \text{sin}\,(2 (\xi_0 + c_1 t)
    +\text{sin}\, (4 (\xi_0 + c_1 t)  \right ]
\lbl{3.43}
\ee
This solution is stable in the sense that the velocity and the kinetic
energy remain finite. The coordinates $\xi$, $\eta$, or equivalently,
$x'$, $y'$, proceed  with time, on average linearly, into infinity.
The velocity thus oscillates around a constant velocity\footnote{
In the quantized theory, to such modulated uniform motion there corresponds
a modulated traveling wave, or uniformly moving wave packet.}.
If we include into the potential also the terms $\frac{1}{2}\omega_1^2 x'^2$ and
$\frac{1}{2}\omega_1^2 y'^2$, then the coordinates do not escape into
infinity, but they oscillate.

\section{Discussion}

It has been shown by some authors\,\ci{Mostafazadeh,Nucci} 
(see also\,\ci{Bolonek}--\ci{Bagarello}) that the Pais-Uhlenbeck oscillator
can be described as a system of two degrees of freedom with positive
definite Hamiltonian. We point out that this holds for the free PU
oscillator only and that one cannot include a couppling term such that
the system would be equivalent to the PU oscillator with a quartic or similar
self-interaction term.
The interacting Pais-Uhlenbeck oscillator must be described, as usually,
by the second order Lagrangian. The Ostrogradski formalism then
leads to the indefinite Hamiltonian, with positive and negative energies.
An equivalent system is that of two oscillators described by the equations
of motion (\ref{3.12}),(\ref{3.13}), derived from the Lagrangian
(\ref{3.11}). We have studied numerical solutions to the latter system for
various coupling constants $\lambda$ and initial velocities. Solutions
are stable below a critical value of $\lambda$ and initial velocity.
We then considered two modifications of the Lagrangian that drastically
increase the range of stability.

Firstly, we replace the quartic interaction term $\frac{\lambda}{4}\,(x'+y')^4$,
that runs into infinity, with the term $\frac{\lambda}{4}\,\text{sin}^4 (x'+y')$
that is finite for all $x'$, $y'$. Then, instead of islands of stability,
we obtain the continent of stability that extends into infinity in the space of
the parameter $\lambda$ and initial conditions. Fig.\,4 shows that now the
system is stable even at $\lambda = 5$, whereas with the quartic interaction
it was unstable already at $\lambda=0.03$. We have done many numerical
runs with higher values of $\lambda$, even with $\lambda=500$, and the
solutions were always stable. By inspecting the
equations of motion, we also found analytically that such interacting
system is indeed stable for any positive $\lambda$, and for any initial
velocity.

Secondly, we replace the kinetic term $\frac{1}{2} ({\dot x}'^2 -{\dot y}'^2)$
with $\frac{1}{2} (m_1 {\dot x}'^2 -m_2 {\dot y}'^2)$, and consider the
case in which the ``masses" $m_1$ and $m_2$ are different. If
$m_1<m_2$, $\lambda >0$, and $\omega_1^2\le\omega_2^2$, the system is stable
for all finite positive values
of $\lambda$ and for all finite positive or negative initial velocities
${\dot x}'(0)$, ${\dot y}'(0)$. Analogously, the system is stable if
$m_1>m_2$, $\lambda <0$, and  $\omega_1^2\ge\omega_2^2$.

Our findings invalidate the generally held belief that the Pais-Uhlenbeck
oscillator in the presence of an interaction is unstable, and therefore
problematic. There are vast regimes of stability that hold for all initial
velocities. This has consequences for the quantum PU oscillator. Namely,
stability of a classical system does not necessarily imply stability of
the corresponding quantum system, because the latter system can tunnel through
a potential barrier and then roll down the potential. But if a classical
system remains stable, regardless of how high is the initial velocity, then
also the quantum system is stable. We conclude that the Pais-Uhlenbeck
oscillator with a suitable self-interaction
 is quite acceptable from the physical point of view. Since
the PU oscillator is a toy model for higher derivative gravity, we expect that
also the negative energy problems of the latter theory could be resolved
along similar lines as investigated in this paper.

\vs{4mm}

\centerline{\bf Acknowledgment}

This work has been supported by the Slovenian Research Agency.


\baselineskip .43cm

\begin{thebibliography}{12}

{\footnotesize

\bi{Ostrogradski} M.V. Ostrogradski, Mem. Acad. Imper. Sci. St. Petersbg., {\bf 6}, 385 (1850).

\bi{Pauli} W. Pauli, Rev. Mod. Phys. {\bf 15}, 175 (1943).

\bi{Jackiw} Y.S. Kim, M.E. Noz, Phys. Rev. D {\bf 8}, 3521 (1973);
{\bf 12} 122 (1975);

\bi{Cangemi}
D. Cangemi, R. Jackiw, B. Zwiebach, Ann. Phys. {\bf 245},408 (1996);

\bi{Benedict}
E. Benedict, R. Jackiw, H.-J. Lee, Phys. Rev. D. {\bf 54}, 6213 (1996).

\bi{PavsicPseudoHarm} M. Pav\v si\v c,
  Phys.\ Lett.\ A {\bf 254}, 119 (1999)
  [hep-th/9812123].

\bibitem{Woodard:2006nt} 
  R.~P.~Woodard,
  Lect.\ Notes Phys.\  {\bf 720}, 403 (2007)
  [astro-ph/0601672].

\bi{Ozonder} S. \"Oz\"onder, "Viable Higher Derivative Theories", PhD Thesis, 2007.

\bi{PU} A. Pais and G.E. Uhlenbeck, Phys. Rev. {\bf 79} 145 (1950).

\bibitem{Smilga1} 
A.~V.~Smilga,
  Phys.\ Lett.\ B {\bf 632}, 433 (2006)
  [hep-th/0503213];
  
\bi{Smilga2}
A.~V.~Smilga,
  Nucl.\ Phys.\ B {\bf 706}, 598 (2005)
  [hep-th/0407231].

\bi{SmilgaStable} A.~V.~Smilga,
  SIGMA {\bf 5}, 017 (2009)
  [arXiv:0808.0139 [quant-ph]].
  
\bi{Robert} D. Robert and A.V. Smilga, J. Math. Phys. {\bf 49}, 042104 (2008).  

\bibitem{Mostafazadeh} 
  A.~Mostafazadeh,
  Phys.\ Lett.\ A {\bf 375}, 93 (2010)
  [arXiv:1008.4678 [hep-th]].

\bi{Bolonek} K. Bolonek, P. Kosi\' nski, Acta Phys. Polon. {\bf 36}, 2115 (2005).

\bi{Sokolov} E.V. Damaskinsky and M.A. Sokolov, J. Phys. A: Math. Gen.
{\bf 39}, 10499 (2006).

\bibitem{Bolonek:2006ir} 
  K.~Bolonek and P.~Kosinski,
  J.\ Phys.\ A {\bf 40}, 11561 (2007)
  [quant-ph/0612091].

\bi{Bagarello} F. Bagarello, Int. J. Theor. Phys. {\bf 50}, 3241 (2011)

\bibitem{Nucci} 
  M.~C.~Nucci and P.~G.~L.~Leach,
  Phys.\ Scripta {\bf 81}, 055003 (2010)
  [arXiv:0810.5772 [math-ph]].

\bibitem{Bender} 
  C.~M.~Bender and P.~D.~Mannheim,
  Phys.\ Rev.\ Lett.\  {\bf 100}, 110402 (2008)
  [arXiv:0706.0207 [hep-th]].

\bibitem{Mannheim} 
  P.~D.~Mannheim,
  Found.\ Phys.\  {\bf 37}, 532 (2007)
  [hep-th/0608154].

\bibitem{Mannheim:2000ka} 
  P.~D.~Mannheim and A.~Davidson,
  ``Fourth order theories without ghosts,''
  hep-th/0001115.

\bibitem{Mannheim:2004qz} 
  P.~D.~Mannheim and A.~Davidson,
  Phys.\ Rev.\ A {\bf 71}, 042110 (2005)
  [hep-th/0408104].
  
\bibitem{Mannheim:2012ci} 
  P.~D.~Mannheim,
  Fortsch.\ Phys.\  {\bf 61}, 140 (2013)
  [arXiv:1205.5717 [hep-th]].

\bibitem{Ilhan} 
  I.~B.~Ilhan and A.~Kovner,
  arXiv:1301.4879 [hep-th].

\bi{PavsicUltrahyper}M.~Pav\v si\v c,
  J.\ Phys.\ Conf.\ Ser.\  {\bf 437}, 012006 (2013)
  [arXiv:1210.6820 [hep-th]].
}

\end{thebibliography}
\end{document}